\documentclass[%
 reprint,
 amsmath,amssymb,
 aps,
]{revtex4-2}

\usepackage{graphicx}
\usepackage{dcolumn}
\usepackage{bm}

\usepackage{graphicx}
\usepackage{dcolumn}
\usepackage{bm}
\usepackage{subcaption}
\usepackage{bm}
\usepackage{amsmath}
\usepackage{siunitx}
\usepackage{soul} 
\usepackage{xcolor}
\usepackage{lineno}

\usepackage{etoolbox}
\usepackage{etoolbox}
\makeatletter
\usepackage{aas_macros}
\makeatother
\definecolor{modblue}{RGB}{255, 255, 204} 
\sethlcolor{modblue} 
\begin{document}

\preprint{APS/123-QED}

\title{Spectral Softenings, Composition Bump, and Anisotropy Transition: A Consistent Picture of Cosmic‑Ray Origin Below the Knee}

\author{Xulin Dong}
\affiliation{College of Physics, Hebei Normal University, No. 20 Road East 2nd Ring South, Shijiazhuang, 050024 Hebei, China}
\affiliation{Key Laboratory of Particle Astrophysics, Institute of High Energy Physics, Chinese Academy of Sciences, No. 19 B Yuquan Road, Shijingshan District, Beijing, 100049, China}
\author{Yuhua Yao} \email{yyao255@wisc.edu}
\affiliation{Wisconsin IceCube Particle Astrophysics Center, University of Wisconsin–Madison, Madison, WI 53703, USA}

\author{Shuwang Cui}\email{cuisw@hebtu.edu.cn}
\affiliation{College of Physics, Hebei Normal University, No. 20 Road East 2nd Ring South, Shijiazhuang, 050024 Hebei, China}
\affiliation{TIANFU Cosmic Ray Research Center, No. 1500 Kezhi Road Chengdu, 610101 Sichuan, China}

\author{Yiqing Guo}
\affiliation{Key Laboratory of Particle Astrophysics, Institute of High Energy Physics, Chinese Academy of Sciences, No. 19 B Yuquan Road, Shijingshan District Beijing 100049, China}
\affiliation{University of Chinese Academy of Sciences, No. 19 A Yuquan Road, Shijingshan District Beijing 100049, China}
\affiliation{TIANFU Cosmic Ray Research Center, No. 1500 Kezhi Road Chengdu, 610101 Sichuan, China}

\date{\today}

\begin{abstract}
Recent DAMPE measurements of individual cosmic-ray components, including carbon, oxygen, and iron, reveal distinct spectral softenings below the knee. The energy spectrum, mass composition, and anisotropy together provide key probes of cosmic-ray origin and propagation. By incorporating the individual elemental spectra reported by DAMPE, we derive a more complete $\langle \ln A\rangle$ distribution, which smoothly connects to the higher-energy $\langle \ln A\rangle$ measurements from LHAASO and exhibits a pronounced bump-like feature. This feature indicates a transition from the conventional Galactic cosmic ray source population to a nearby-source-dominated regime. We show that a spatially dependent propagation model with a nearby-source contribution can consistently reproduce the observed spectra, mass composition, and anisotropy. This suggests a unified picture in which Galactic cosmic rays below the knee arise from multiple source populations jointly constrained by these observables. Leveraging the precise component-resolved spectra from DAMPE, we further predict the transition energies in the anisotropy phase and amplitude for different mass components. Future component-resolved anisotropy measurements by LHAASO will provide a crucial test of this scenario.


\end{abstract}

\maketitle

\section{\label{sec:level1}Introduction}
The origin of Galactic cosmic rays (CRs) has remained an open question for more than a century (see \cite{2020PhR...872....1B} for a review). Observational data on the CR energy spectrum, mass composition, and arrival-direction anisotropy provide key probes of CR acceleration, propagation, and interactions. Understanding these observables near Earth represents a milestone toward the identification of their origin. In the conventional picture of Galactic CRs \citep{1983RPPh...46..973D,2007ARNPS..57..285S}, particles are accelerated by astrophysical shocks and subsequently diffuse through the turbulent Galactic magnetic field in a rigidity-dependent manner, producing approximately power-law spectra, a rigidity-dependent evolution of the mass composition, and a small large-scale anisotropy. However, recent high-precision measurements have revealed multiple deviations from this simple picture, challenging conventional expectations and indicating that additional source, acceleration, or propagation effects may be required.

Great progress in CR spectral and composition measurements has been achieved through direct space-borne and balloon-borne experiments, as well as ground-based air-shower observations. Direct measurements by ATIC-2, CREAM, and PAMELA revealed deviations from simple power-law spectra and species-dependent behavior of the proton and helium components \citep{2009BRASP..73..564P,2011Sci...332...69A,2011ApJ...728..122Y}. High-precision AMS-02, DAMPE, and CALET measurements further showed similar hardenings of primary nuclei above $\sim200$~GV and stronger hardenings of secondary nuclei \citep{2017PhRvL.119y1101A,2018PhRvL.120b1101A,2019SciA....5.3793A,2021PhRvL.126t1102A,2019PhRvL.122r1102A,2020PhRvL.125y1102A,2022PhRvL.129j1102A,2023PhRvL.130q1002A}. The interpretation of these features involves source-injection effects or source superposition \citep{2011PhRvD..84d3002Y,2019SCPMA..6249511Y}, nearby-source contributions \citep{2012MNRAS.421.1209T,pzxy-v9v8,Yue_2019}, nonlinear propagation driven by CR-induced turbulence \citep{2012PhRvL.109f1101B,Yuan_2017}, and spatially dependent diffusion \citep{2012ApJ...752L..13T,2016ApJ...819...54G}.

At higher energies, DAMPE and CALET have reported softenings in the proton and helium spectra at tens of TeV energies, while recent DAMPE measurements of C, O, and Fe indicate a common softening rigidity of about $15$~TV for heavier primary nuclei \citep{2019SciA....5.3793A,2021PhRvL.126t1102A,2024PhRvD.109l1101A,2026Natur.653...52D}. Ground-based measurements by LHAASO and GRAPES-3 further reveal structures in the light components, including proton hardening around $166$~TeV, a proton knee near a few PeV, and helium hardening followed by softening at PeV energies \citep{2024PhRvL.132e1002V,2025SciBu..70.4173C,2026PhRvL.136l1001C}. Together, these results suggest multiple rigidity-dependent structures beyond a single smooth power-law description.

These spectral structures are closely connected to the evolution of the CR mass composition. Recent component-resolved DAMPE measurements, especially for C, O, and Fe, enable a more complete data-driven estimate of $\langle \ln A\rangle$ below the knee \citep{2026Natur.653...52D}. In the knee region, precise LHAASO measurements further suggest that the all-particle knee is mainly associated with the light component rather than medium-heavy nuclei \citep{2024PhRvL.132m1002C}. Together, DAMPE and LHAASO provide continuous composition constraints from direct measurements to the air-shower regime.


While the energy spectrum and mass composition encode the energy-dependent behavior of different nuclear species, the arrival-direction anisotropy provides complementary information on the spatial distribution of CR sources and the properties of Galactic transport \citep{2016PhRvL.117o1103A}. In simple diffusion models, the anisotropy phase is expected to be broadly related to the large-scale Galactic CR density gradient, which may be connected to the distribution of Galactic sources toward the inner Galaxy \citep{2012JCAP...01..011B}. However, observations in the TeV--PeV range reveal a more complex picture, with a small relative amplitude, strong energy dependence, and both large- and small-angular-scale structures \citep{2009ApJ...698.2121A,2012ApJ...746...33A,2015ApJ...809...90B,2017ApJ...836..153A,2017PrPNP..94..184A}. Recent LHAASO-KM2A measurements further suggest a composition dependence of the anisotropy: a high-purity proton sample in the 10--220~TeV range shows an amplitude and phase turnover at lower energy than heavier-nuclei samples \citep{LHAASO2025ProtonAnisotropy}. These results indicate that CR anisotropy is shaped not only by the global source distribution, but also by the interplay between composition, local sources, and Galactic magnetic-field structure.

The joint interpretation of the CR energy spectrum, mass composition, and arrival-direction anisotropy has become an important approach for studying the origin of ultra-high-energy (E $\ge10^{18}$ eV) CRs \citep{2019FrASS...6...23B,2024JCAP...01..022A,qiao2026coevolutioncosmicrayenergy}. Extending this strategy to Galactic CRs at lower energies is equally valuable, but has been more limited because the sub-PeV to PeV region lies between the high-energy reach of direct space-borne measurements and the low-energy threshold of ground-based air-shower observations. Recent measurements from DAMPE and LHAASO now provide improved constraints across this transition region, making it possible to connect direct component-resolved spectra with indirect composition and anisotropy measurements.

Previous theoretical and phenomenological studies have commonly interpreted the Galactic CR energy spectrum, mass composition, and anisotropy either separately or through pairwise combinations. Spatially dependent propagation models have been invoked to explain the observed spectral hardenings \citep{2012ApJ...752L..13T,2016ApJ...819...54G,PhysRevD.97.063008}, while the inclusion of a nearby source can help reproduce the energy-dependent anisotropy amplitude and phase \citep{2012JCAP...01..011B,2016PhRvL.117o1103A,2023ApJ...956...75Q,2026ApJ...996...77Q}. In this work, we jointly analyze three complementary probes—the component-resolved energy spectrum, the mean logarithmic mass, and the large-scale anisotropy—to examine whether a spatially dependent propagation model with a nearby-source contribution can provide a unified description of Galactic CR observations. Section~II introduces the model and methodology used in this work, Section~III presents the results and discussion, and Section~IV summarizes our conclusions.

\section{MODEL AND Method}


\begin{table*}[!htb]
\caption{\label{tab:table2}Spectral injection parameters.}
\begin{ruledtabular}
\begin{tabular}{ccccccccccc}
 &&Type-A&&Nearby source&&&Type-B&\\
 & Normalization &$\quad\gamma_2$& Abundance& $q_0$ &$\quad\gamma$&Normalization &$\quad\gamma_2$\footnotemark[1]& Abundance&\\
&$[\mathrm{GeV}^{-1}\mathrm{m}^{-2}\mathrm{s}^{-1}\mathrm{sr}^{-1}]$&&&$[\mathrm{GeV}^{-1}]$&&$[\mathrm{GeV}^{-1}\mathrm{m}^{-2}\mathrm{s}^{-1}\mathrm{sr}^{-1}]$ \\ \hline
H&$4.10\times 10^{-2}$&$2.42$&$1.10\times10^{6}$&$1.3\times10^{52}$&2&$4.00\times 10^{-3}$&$2.07$&$1.10\times10^{6}$ \\
He&$2.38\times 10^{-3}$&$2.34$&71300&$2.2\times10^{51}$&2&$5.09\times 10^{-4}$&$2.07$&$1.4\times10^{5}$ \\
C&$8.84\times 10^{-5}$&$2.36$&2650&$3.5\times10^{49}$&2&$9.64\times 10^{-6}$&$2.07$&$2650$ \\
O&$1.12\times 10^{-4}$&$2.37$&3350&$2.0\times10^{49}$&2&$1.22\times 10^{-5}$&$2.07$&$3350$ \\
Ne&$1.97\times 10^{-5}$&$2.38$&590&$4.0\times10^{48}$&2&$9.64\times 10^{-7}$&$2.07$&$265$ \\
Mg&$2.37\times 10^{-5}$&$2.40$&710&$2.2\times10^{48}$&2&$1.16\times 10^{-6}$&$2.07$&$320$ \\
Si&$2.29\times 10^{-5}$&$2.40$&690&$2.0\times10^{48}$&2&$1.13\times 10^{-6}$&$2.07$&$310$ \\
Fe&$2.38\times 10^{-5}$&$2.42$&700&$1.6\times10^{48}$&2&$1.15\times 10^{-6}$&$2.07$&$315$ \\
\end{tabular}
\end{ruledtabular}
\footnotetext[1]{$\gamma_1=2$, $R_{br}=7.2$~GV. The cutoff magnetic rigidity of Type-A sources is $R_{\rm cut,A}=100$~TV, that of Type-B sources is $R_{\rm cut,B}=4$~PV, and that of the nearby source is $R_{\rm cut,nearby}=25$~TV.}
\end{table*}

\begin{table}[]
\caption{\label{tab:table1}Propagation model parameters.}
\begin{ruledtabular}
\begin{tabular}{cccccccccccccccc}
$D_0$ &$\delta_0$&$\quad N_m$&$\quad\xi$&$\quad n$&$\quad v_a$\footnotemark[1]&$\quad z_0$&$r_0$\ \\
$[\mathrm{cm}^{2}\mathrm{s}^{-1}]$&&&&&$[\mathrm{km~s}^{-1}]$&$[\mathrm{kpc}]$&$[\mathrm{kpc}]$\\ \hline
 $4.87\times10^{28}$&0.58&0.52&0.10&4.00&6.00&5.00&20.00 \\
\end{tabular}
\end{ruledtabular}
\footnotetext[1]{$v_a$ is the Alfvén velocity.}
\end{table}

The local CR observables measured near Earth are shaped by the combined effects of particle acceleration at astrophysical source sites, subsequent transport through the interstellar medium, and interactions during propagation. This transport process is commonly described by the CR propagation equation \cite{maurin2002galactic}:
\begin{equation}
\begin{split}
\frac{\partial\Psi(\bm{r},p,t)}{\partial t} = Q(\vec{r},p,t)+\vec{\nabla}\cdot(D_{xx}\vec{\nabla}\Psi-\vec{V}_{c}\Psi) \\
+ \frac{\partial}{\partial p}[p^{2}D_{pp}\frac{\partial}{\partial p}\frac{\Psi}{p^{2}}]
-\frac{\partial}{\partial p}[\dot{p}\Psi-\frac{p}{3}(\vec{\nabla}\cdot\vec{V}_{c})\Psi]\\
-\frac{\Psi}{\tau_{f}}-\frac{\Psi}{\tau_{r}},
\end{split}
\end{equation}

Here, $\Psi(\vec{r},p,t)$ is the CR density in phase space, and $Q(\vec{r},p,t)$ represents the source injection term. The remaining terms describe the main transport and loss processes: spatial diffusion through $D_{xx}$, convection with velocity $\vec{V}_c$, diffusive reacceleration described by the momentum diffusion coefficient $D_{pp}$, continuous energy losses represented by $\dot{p}$, fragmentation with timescale $\tau_f$, and radioactive decay with timescale $\tau_r$. In the following subsections, we provide a more detailed description of the source injection and spatial diffusion terms.

\subsection{Source Injection}
The source term $Q(\vec{r},p,t)$ describes where CRs are injected in the Galaxy and with what spectrum. In this work, we consider three source components: Type-A sources, Type-B sources, and a nearby source. Type-A sources represent the conventional Galactic CR population, with acceleration limits of several hundred TeV per nucleon, as expected for typical supernova remnants. Type-B sources describe a higher-energy Galactic population capable of accelerating particles up to several PeV per nucleon. Recent LHAASO observations suggest that sources such as microquasars may reach this energy range \citep{2025NSRev..12af496L,2025arXiv251216638T}. In addition, a nearby source is introduced to account for local spectral features and the observed large-scale anisotropy. The Type-A and Type-B source populations are treated as steady-state continuous source distributions, while the nearby source is described as an impulsive injection event with a finite source age.

In the steady-state approximation, the source term of each Galactic source population can be factorized into a spatial distribution and an injection spectrum,
\begin{equation}
Q_i(\vec{r},p)=f_i(r,z)q(R),
\end{equation}
where $i=A,B$, $f_i(r,z)$ gives the Galactic spatial distribution of the corresponding source population, and $q(R)$ describes its rigidity-dependent injection spectrum. The spatial distribution of Type-A sources is approximated by that of supernova remnants \citep{1996A&AS..120C.437C},
\begin{equation}
f_{A}(r,z)=
\left(\frac{r}{r_\odot}\right)^{1.25}
\exp\left[-\frac{3.87(r-r_\odot)}{r_\odot}\right]
\exp\left(-\frac{|z|}{z_s}\right),
\end{equation}
where $r_{\odot}=\SI{8.5}{kpc}$ and $z_s=\SI{0.2}{kpc}$. The spatial distribution of Type-B sources is approximated by that of X-ray binaries \citep{2026ApJ...997..163Y},
\begin{equation}
f_{B}(r,z)=
\frac{
\exp \left(-a\frac{r}{R_{\odot}}\right)
+B\exp \left(-b\frac{r^2}{R_{\odot}^2}\right)
}{
\exp(-a)+B\exp(-b)
}
\exp\left(-\frac{|z|}{z_s}\right),
\end{equation}
where $a=2.67\pm0.21$, $b=39.33\pm3.88$, and $B=9.74\pm1.99$.

The injection spectra of both Type-A and Type-B sources are assumed to follow broken power laws in rigidity with exponential cutoffs,
\begin{equation}
q(R) \propto
\begin{cases}
\left(\frac{R}{R_{\rm br}}\right)^{-\gamma_1},
& R < R_{\rm br}, \\
\left(\frac{R}{R_{\rm br}}\right)^{-\gamma_2}
\exp\left[-\frac{R}{R_c}\right],
& R \ge R_{\rm br}.
\end{cases}
\end{equation}
Here, $R_{\rm br}$ is the break rigidity, $R_c$ is the cutoff rigidity, and $\gamma_1$ and $\gamma_2$ are the injection spectral indices below and above the break, respectively.

To reproduce both the observed energy spectra and anisotropy, we further include a nearby source located approximately in the anti-Galactic-center direction. This source is modeled as an instantaneous injection event,
\begin{equation}
Q(R,t)=q_0\delta(t-t_0)
\left(\frac{R}{R_0}\right)^{-\gamma}
\exp\left[-\frac{R}{R_c}\right],
\end{equation}
where $t_0$ denotes the effective time of CR injection by the nearby source, $q_0$ is the normalization, $\gamma$ is the injection spectral index, and $R_c$ is the cutoff rigidity. Following previous studies \citep{2026ApJ...996...77Q}, we assume a source location of $(\mathrm{R.A.}=4^{\rm h}0^{\rm m}, \delta=24^\circ30')$, a distance of about $0.3~\mathrm{kpc}$, and an age of $3\times10^5~\mathrm{yr}$, close to the Geminga region. The normalization $q_0$ is determined by fitting the spectra of various elements observed by AMS-02 and DAMPE. The detailed injection parameters are listed in Table~I.

\begin{figure*}[!htb]  
    \centering
    \begin{subfigure}[t]{0.23\textwidth}  
        \centering
        \includegraphics[
            width=\linewidth,
            height=3.5cm,                
            keepaspectratio,             
            bb=0 0 500 400,              
            draft=false                  
        ]{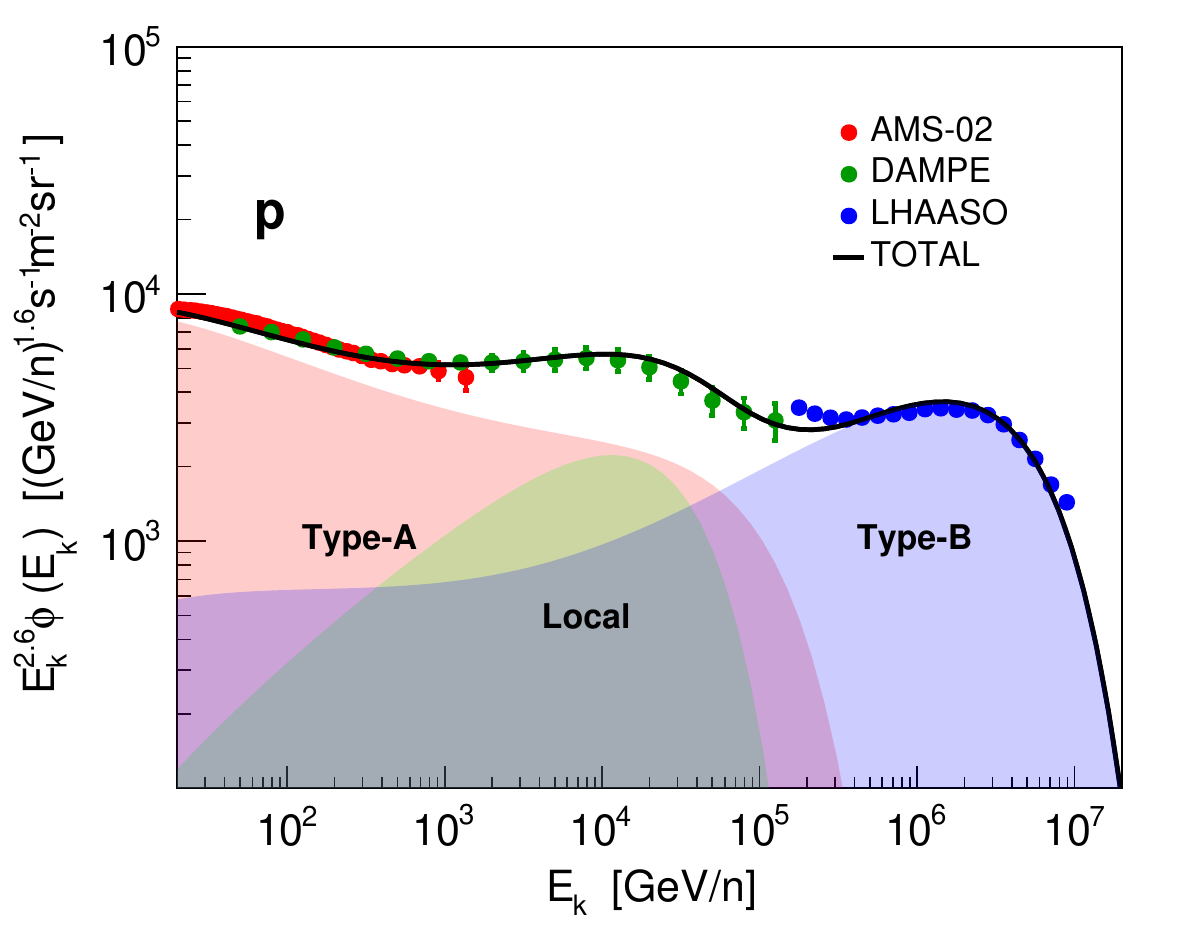}
        \label{fig:he_ams}
    \end{subfigure}
    \hfill
    \begin{subfigure}[t]{0.23\textwidth}
        \centering
        \includegraphics[
            width=\linewidth,
            height=3.3cm,
            bb=0 0 500 400,
            draft=false
        ]{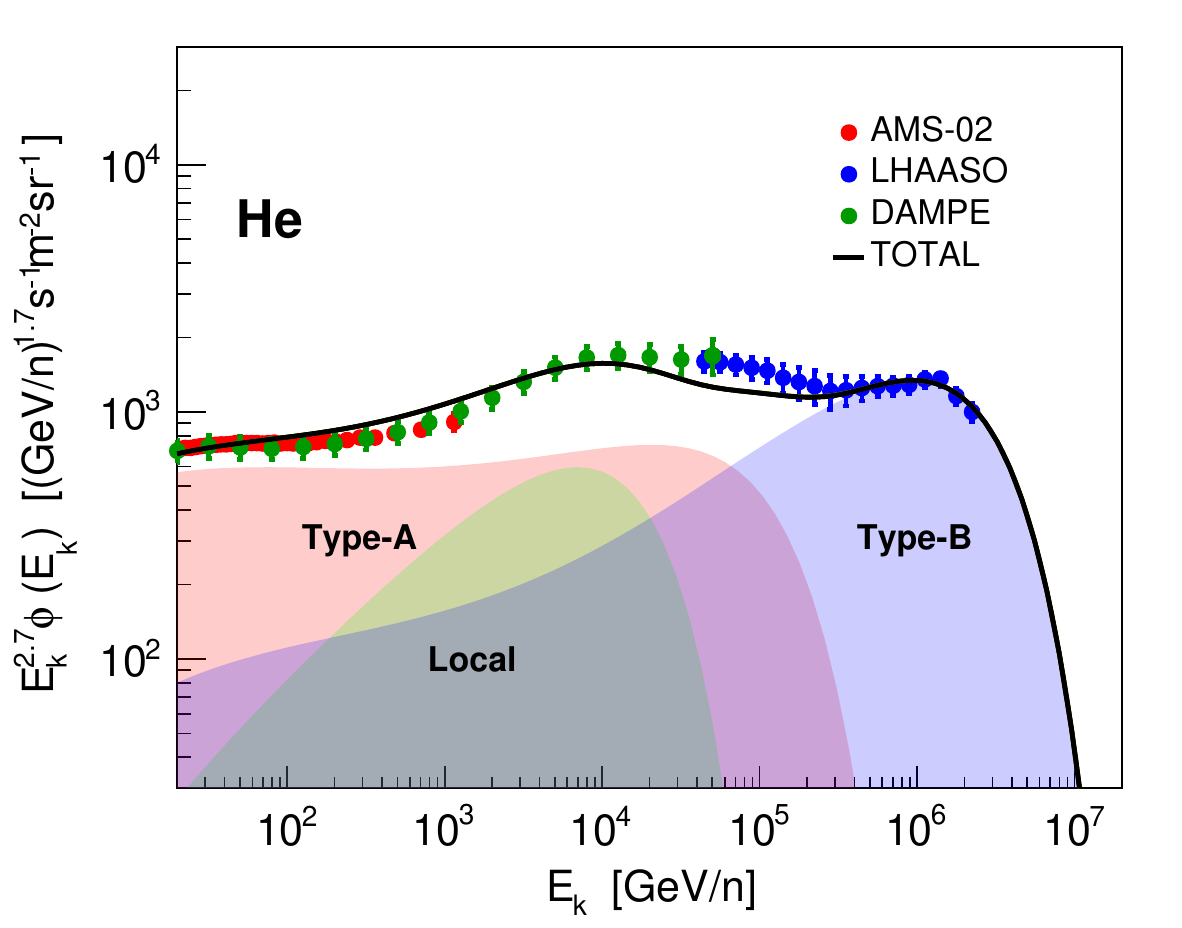}
        \label{fig:b_ams}
    \end{subfigure}
    \hfill
    \begin{subfigure}[t]{0.23\textwidth}
        \centering
        \includegraphics[
            width=\linewidth,
            height=3.3cm,
            bb=0 0 500 400,
            draft=false
        ]{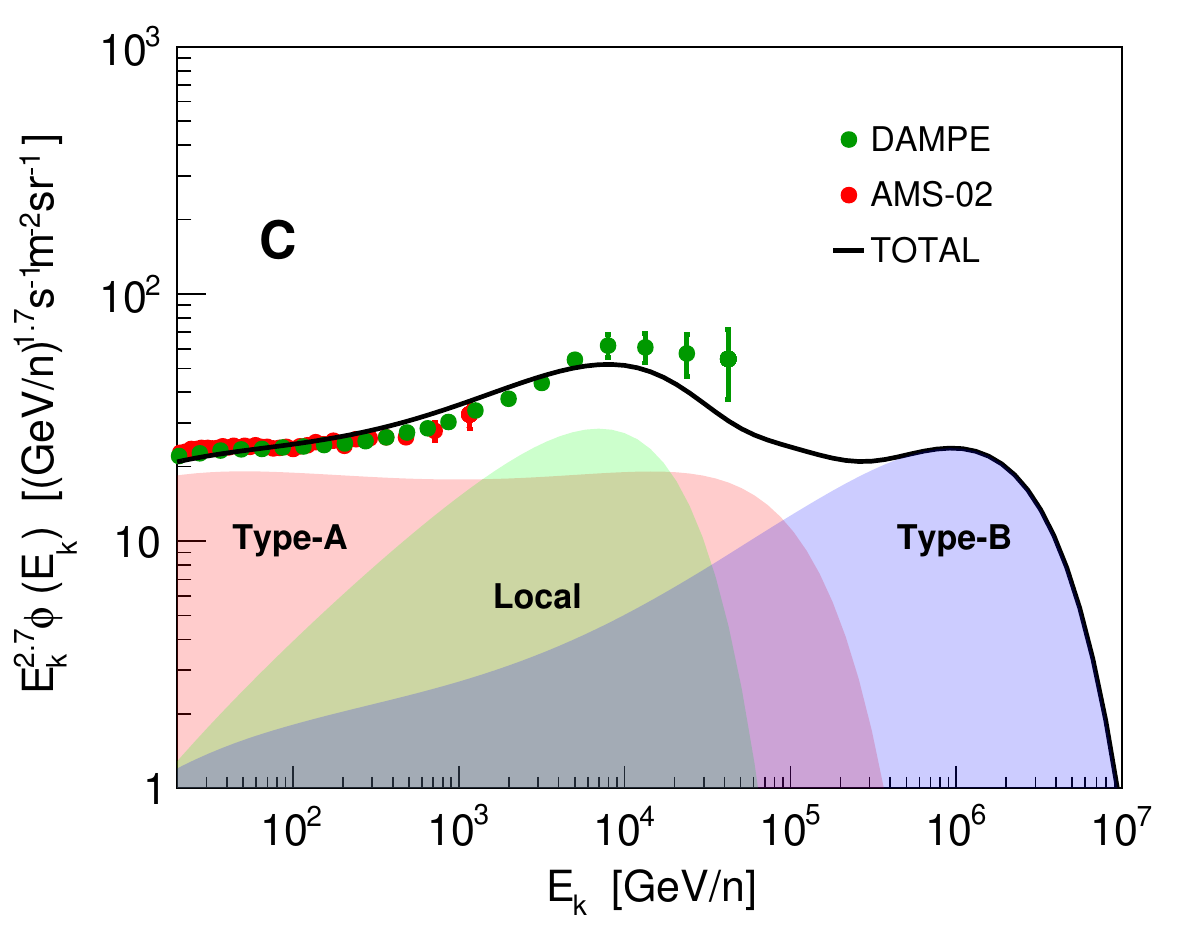}
        \label{fig:c_ams}
    \end{subfigure}
    \hfill
    \begin{subfigure}[t]{0.23\textwidth}
        \centering
        \includegraphics[
            width=\linewidth,
            height=3.3cm,
            bb=0 0 500 400,
            draft=false
        ]{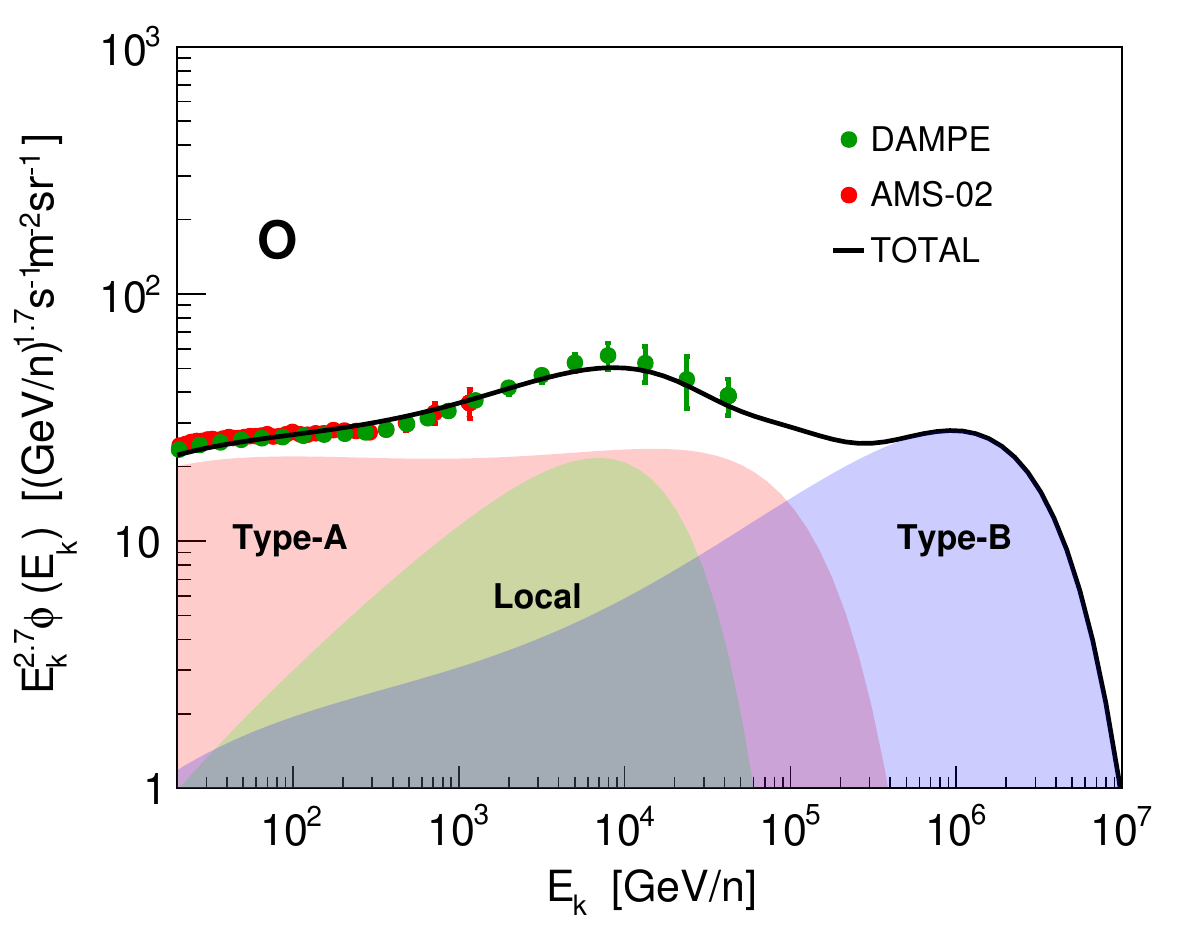}
        \label{fig:n_ams}
    \end{subfigure}

    \vspace{0.2cm}

    \begin{subfigure}[t]{0.23\textwidth}
        \centering
        \includegraphics[
            width=\linewidth,
            height=3.3cm,
            bb=0 0 500 400,
            draft=false
        ]{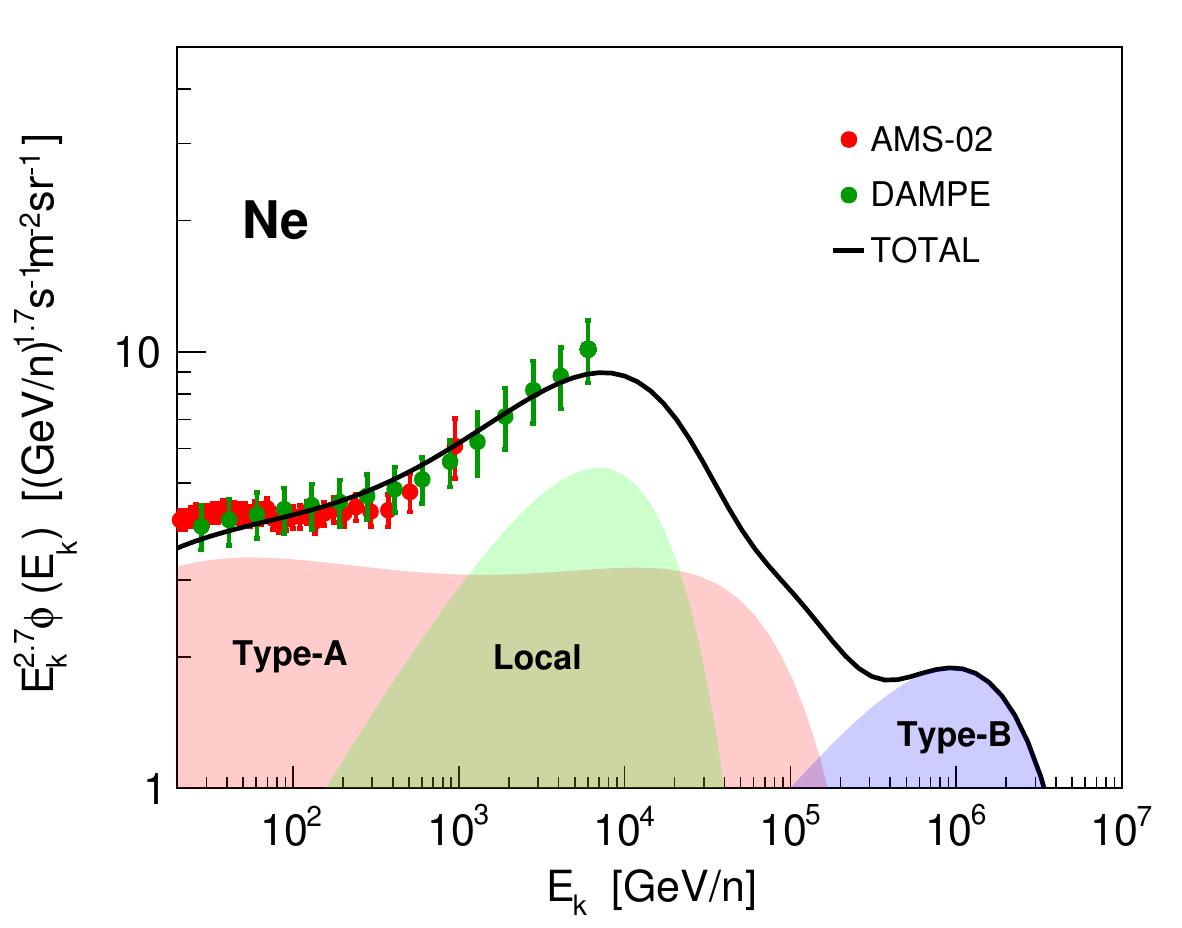}
        \label{fig:o_ams}
    \end{subfigure}
    \hfill
    \begin{subfigure}[t]{0.23\textwidth}
        \centering
        \includegraphics[
            width=\linewidth,
            height=3.3cm,
            bb=0 0 500 400,
            draft=false
        ]{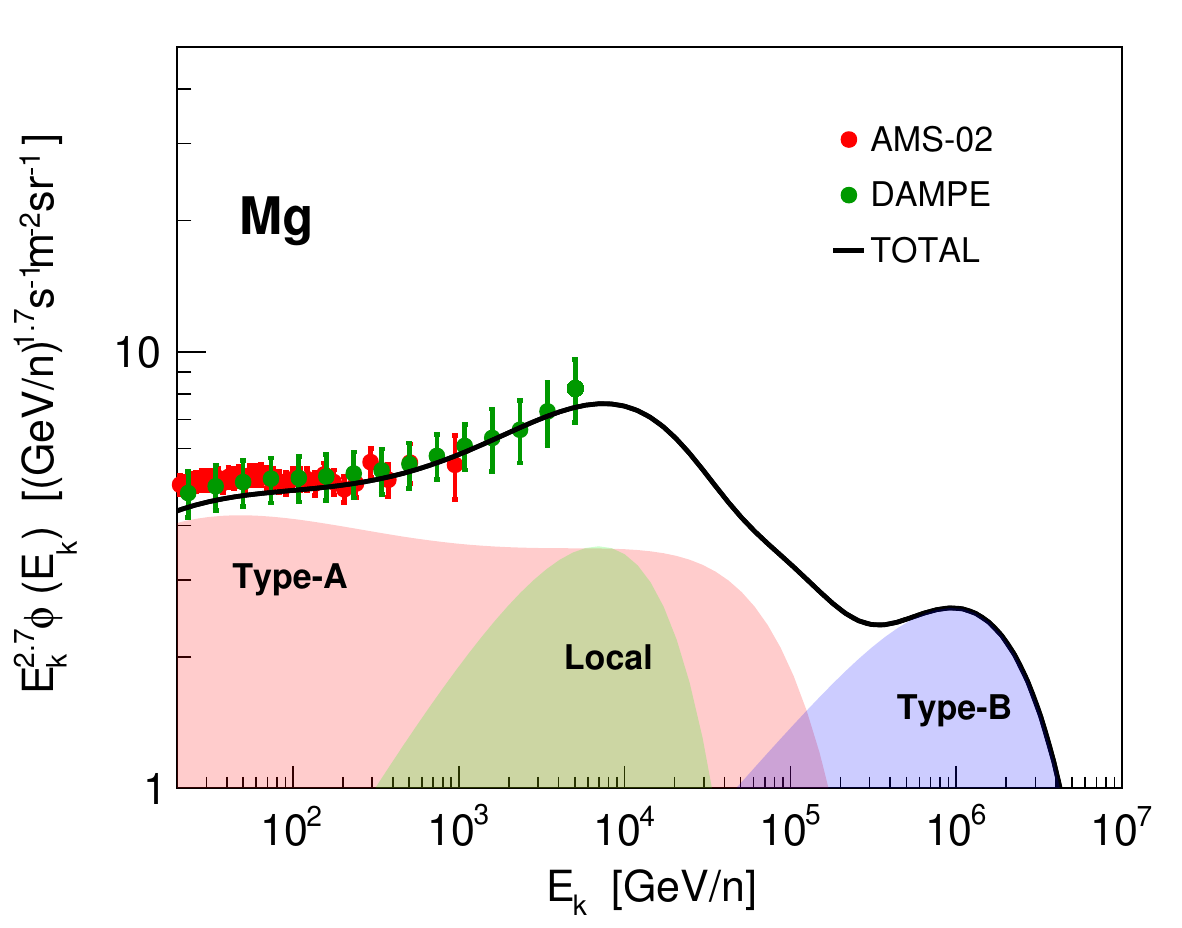}
        \label{fig:ne_ams}
    \end{subfigure}
    \hfill
    \begin{subfigure}[t]{0.23\textwidth}
        \centering
        \includegraphics[
            width=\linewidth,
            height=3.3cm,
            bb=0 0 500 400,
            draft=false
        ]{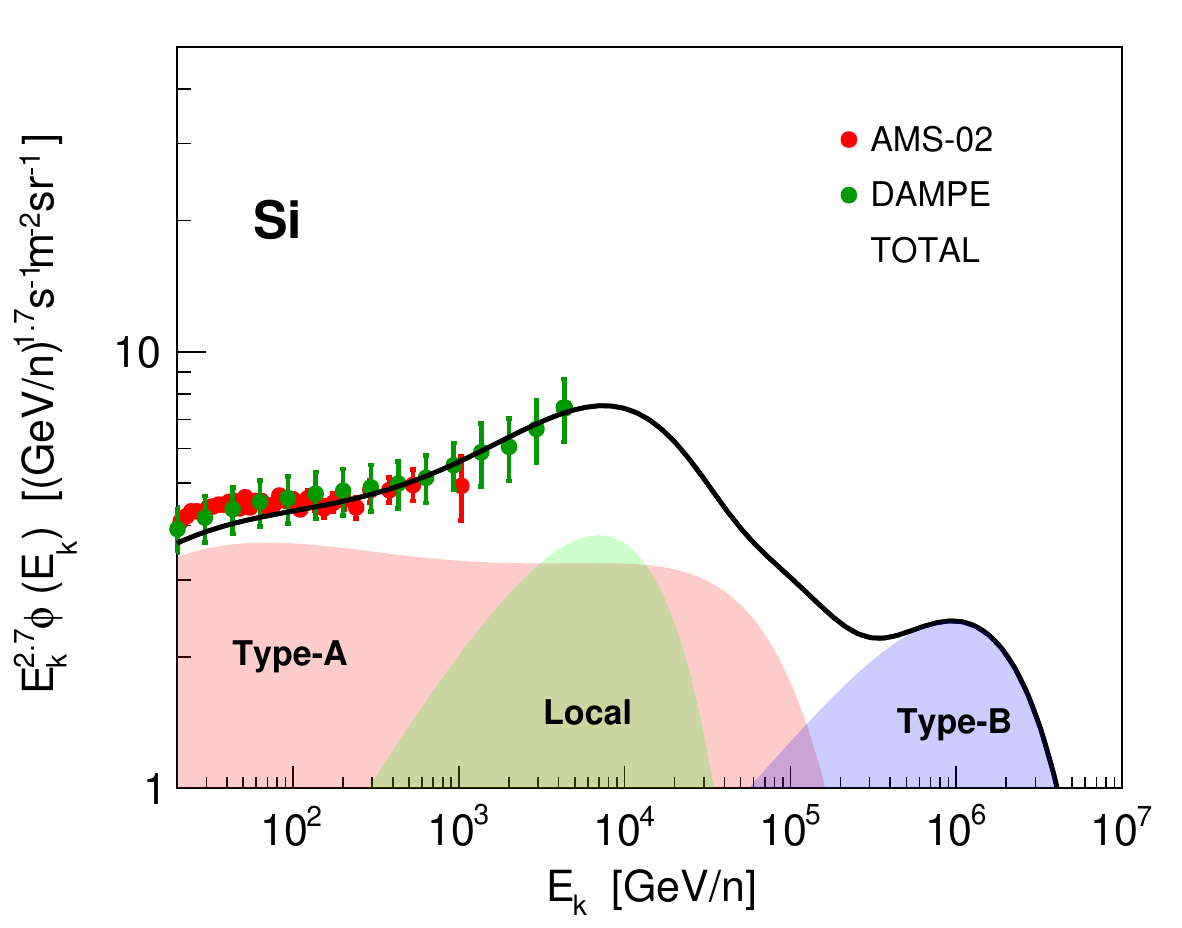}
        \label{fig:na_ams}
    \end{subfigure}
    \hfill
    \begin{subfigure}[t]{0.23\textwidth}
        \centering
        \includegraphics[
            width=\linewidth,
            height=3.3cm,
            bb=0 0 500 400,
            draft=false
        ]{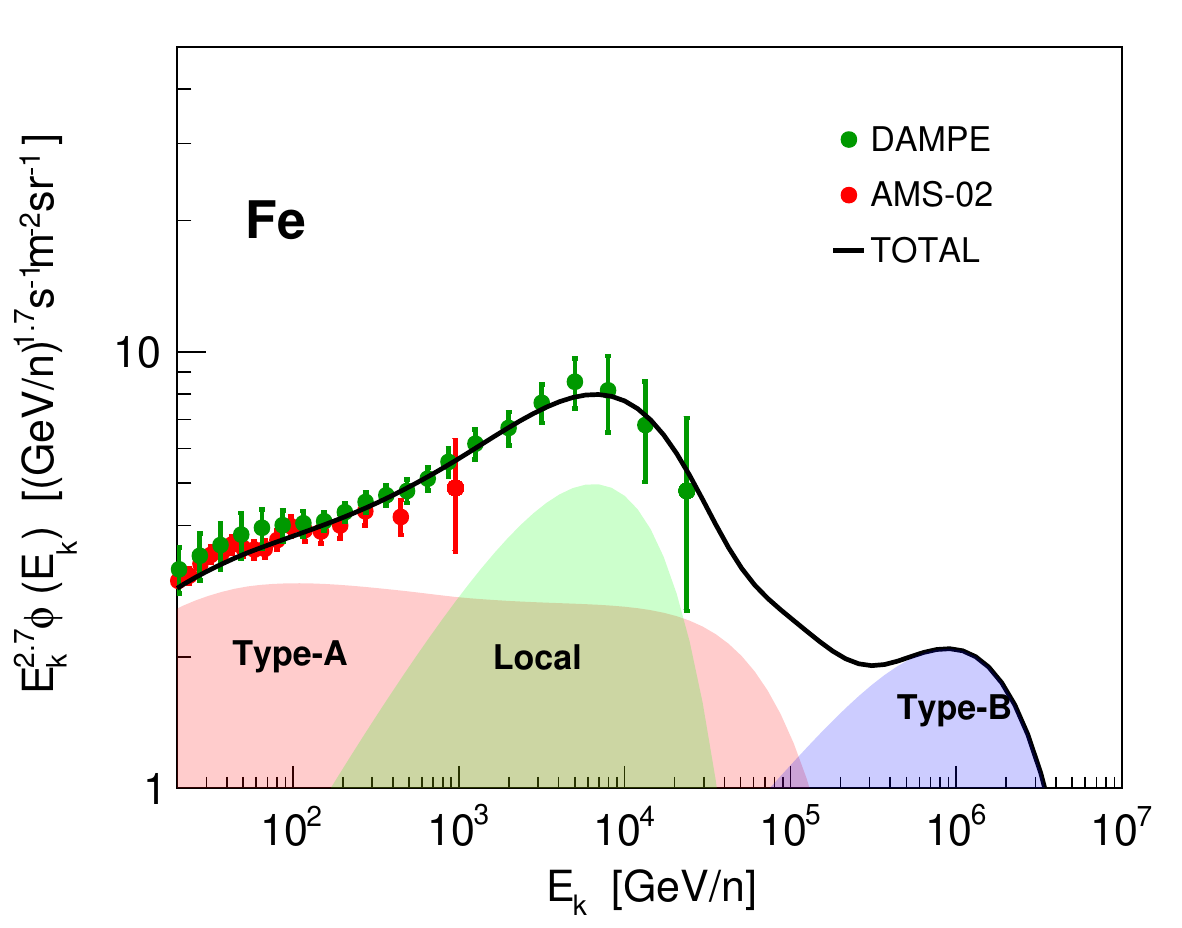}
        \label{fig:mg_ams}
    \end{subfigure}

    \caption{
Model-predicted energy spectra of individual CR nuclei compared with measurements from AMS-02 \cite{aguilar2021alpha,aguilar2024properties,aguilar2023properties,aguilar2021properties1,aguilar2020properties,aguilar2019properties}, DAMPE \cite{2026Natur.653...52D,Casilli:2025w8}, and LHAASO \cite{2025SciBu..70.4173C,2026PhRvL.136l1001C}.
In the figure, red, blue, and green points represent AMS-02, DAMPE, and LHAASO data, respectively; the black solid line shows the total model flux. The labels "Local", "Type-A", and “Type-B” denote the energy spectra of the nearby source, Type-A sources and Type-B sources, respectively.
}
    \label{fig:element_spectra_35cm}
\end{figure*}

\begin{figure}
    \centering
    \includegraphics[width=1\linewidth]{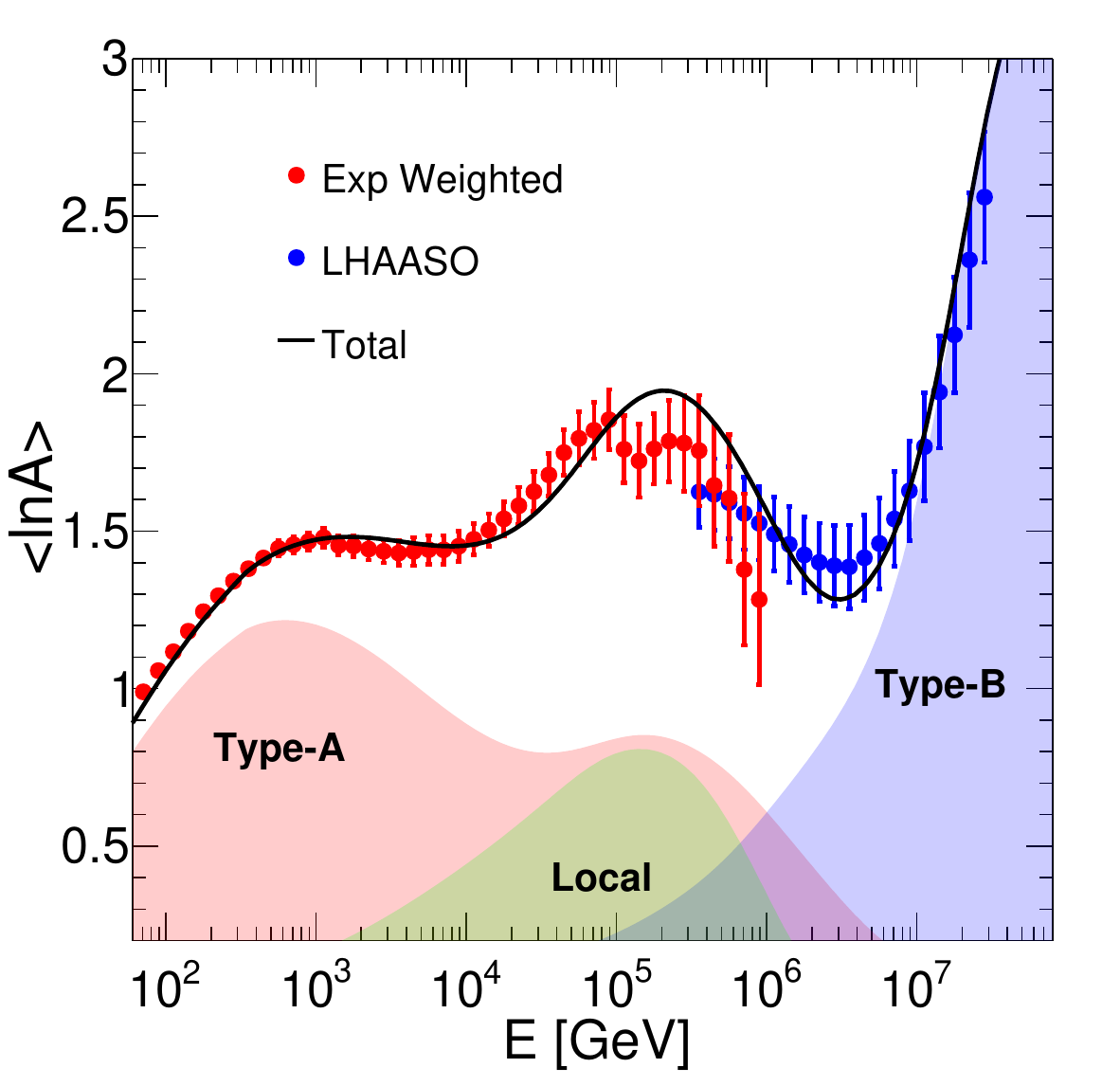}
    \caption{Comparison between the $\langle \ln A\rangle$ calculated by the model and the $\langle \ln A\rangle$ calculated from experimental data. The red points represent the $\langle \ln A\rangle$ calculated from the AMS-02 and DAMPE experimental data, the blue points represent the $\langle \ln A\rangle$ measured by the LHAASO experiment, and the black solid line represents the $\langle \ln A\rangle$ calculated by the model. The shaded regions labeled "Type-A", "Type-B", and "Local" represent the contributions of the three types of sources to the total $\langle \ln A\rangle$, respectively.}
    \label{fig:placeholder}
\end{figure}

\begin{figure*}[]  
    \centering
    \begin{subfigure}[]{0.48\textwidth}  
        \centering
        \includegraphics[
            width=\linewidth,
            height=7cm,                  
            keepaspectratio,             
            draft=false                  
        ]{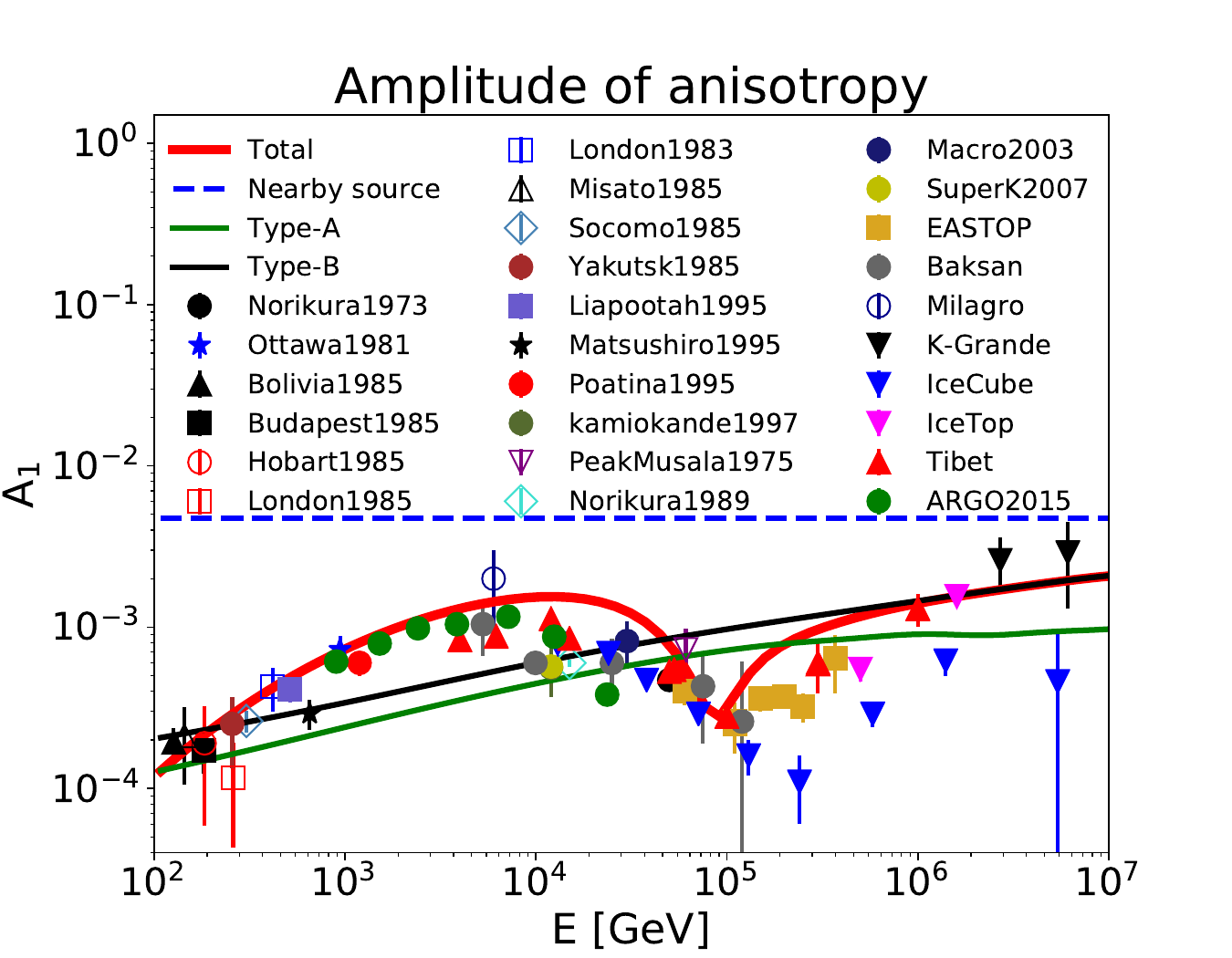}
        \label{fig:inj_ams}
    \end{subfigure}
    \hfill
    \begin{subfigure}[]{0.48\textwidth}
        \centering
        \includegraphics[
            width=\linewidth,
            height=7cm,                  
        ]{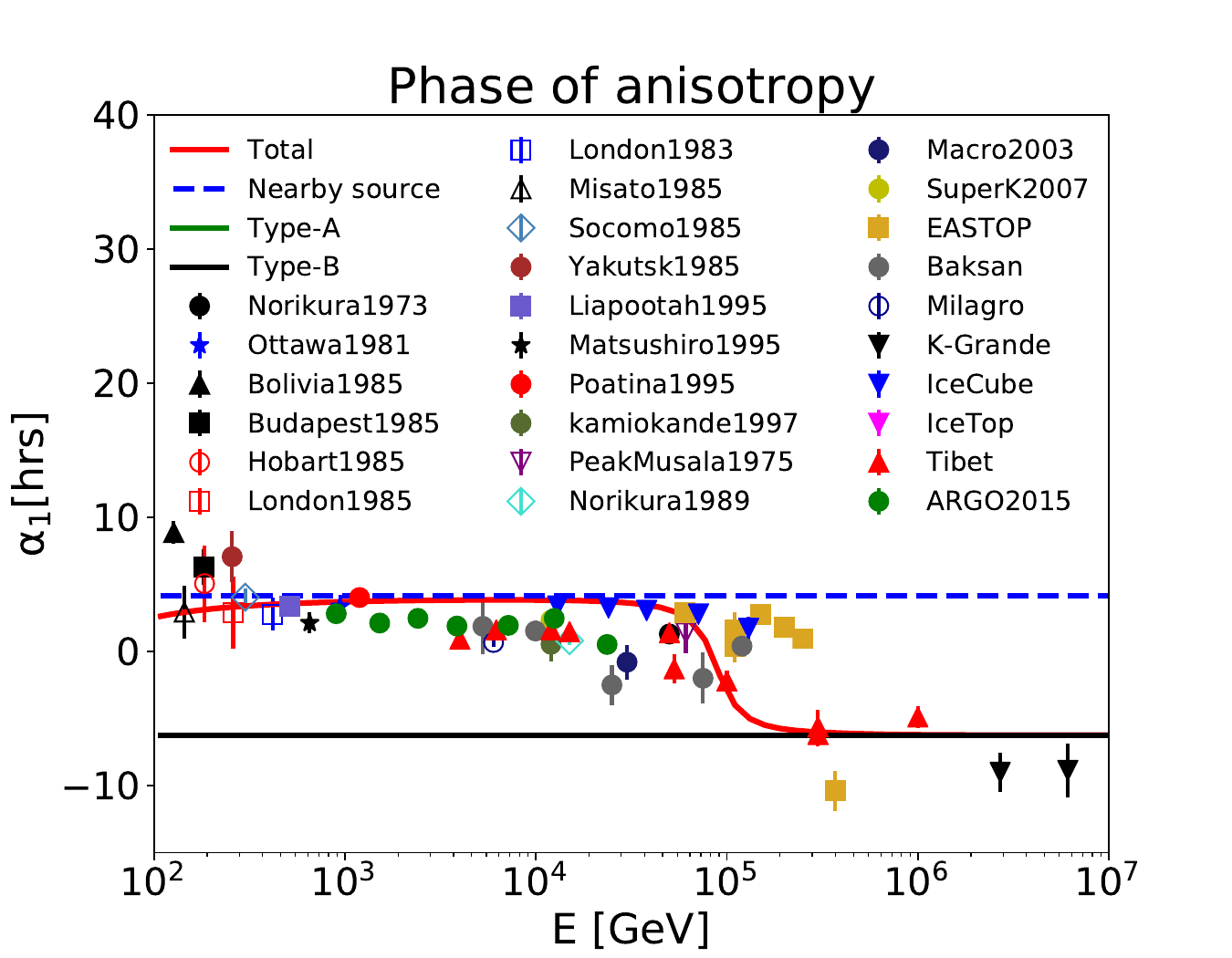}
        \label{fig:inj_dampe}
    \end{subfigure}

    \caption{The amplitude (left) and phase (right) of CR anisotropy. The solid red line shows the total anisotropy from the Type-A, Type-B, and nearby-source components. The dashed and solid colored lines show the individual component contributions. The data points are taken from Norikura \cite{1973ICRC....2.1058S}, Ottawa\cite{1981ICRC...10..246B}, London\cite{1983ICRC....3..383T}, Bolivia\cite{1985P&SS...33.1069S}, Budapest\cite{1985P&SS...33.1069S}, Hobart\cite{1985P&SS...33.1069S}, London\cite{1985P&SS...33.1069S}, Misato\citep{1985P&SS...33.1069S}, Socorro\cite{1985P&SS...33.1069S}, Yakutsk\cite{1985P&SS...33.1069S}, Banksan\cite{1987ICRC....2...22A}, HongKong\cite{1987ICRC....2...18L}, Sakashita\cite{1990ICRC....6..361U}, Utah\cite{1991ApJ...376..322C}, Liapootah\cite{1995ICRC....4..648M}, Matsushiro\cite{1995ICRC....4..639M}, Poatina\cite{1995ICRC....4..635F},
    Kamiokande\cite{1997PhRvD..56...23M}, Marco\cite{2003PhRvD..67d2002A}, SuperKamiokande\cite{2007PhRvD..75f2003G}, PeakMusala\cite{1975ICRC....2..586G}, Baksan\cite{1981ICRC....2..146A}, Norikura\cite{1989NCimC..12..695N}, EAS-TOP\cite{1995ICRC....2..800A,1996ApJ...470..501A,2009ApJ...692L.130A}, Baksan\cite{alekseenko200910}, Milagro\cite{2009ApJ...698.2121A}, KASCADE-Grande\cite{2015ICRC...34..281C}, IceCube\cite{2010ApJ...718L.194A,2012ApJ...746...33A}, Ice-Top\cite{2013ApJ...765...55A}, ARGOYBJ\cite{2015ApJ...809...90B}, Tibet\cite{2005ApJ...626L..29A,2017ApJ...836..153A,2015ICRC...34..355A}, and AUGER\cite{2020ApJ...891..142A}.
    }
    \label{fig:injection_spectra}
\end{figure*}
\begin{figure*}[]  
    \centering
    \begin{subfigure}[]{0.48\textwidth}  
        \centering
        \includegraphics[
            width=\linewidth,
            height=7cm,                  
            keepaspectratio,             
            draft=false                  
        ]{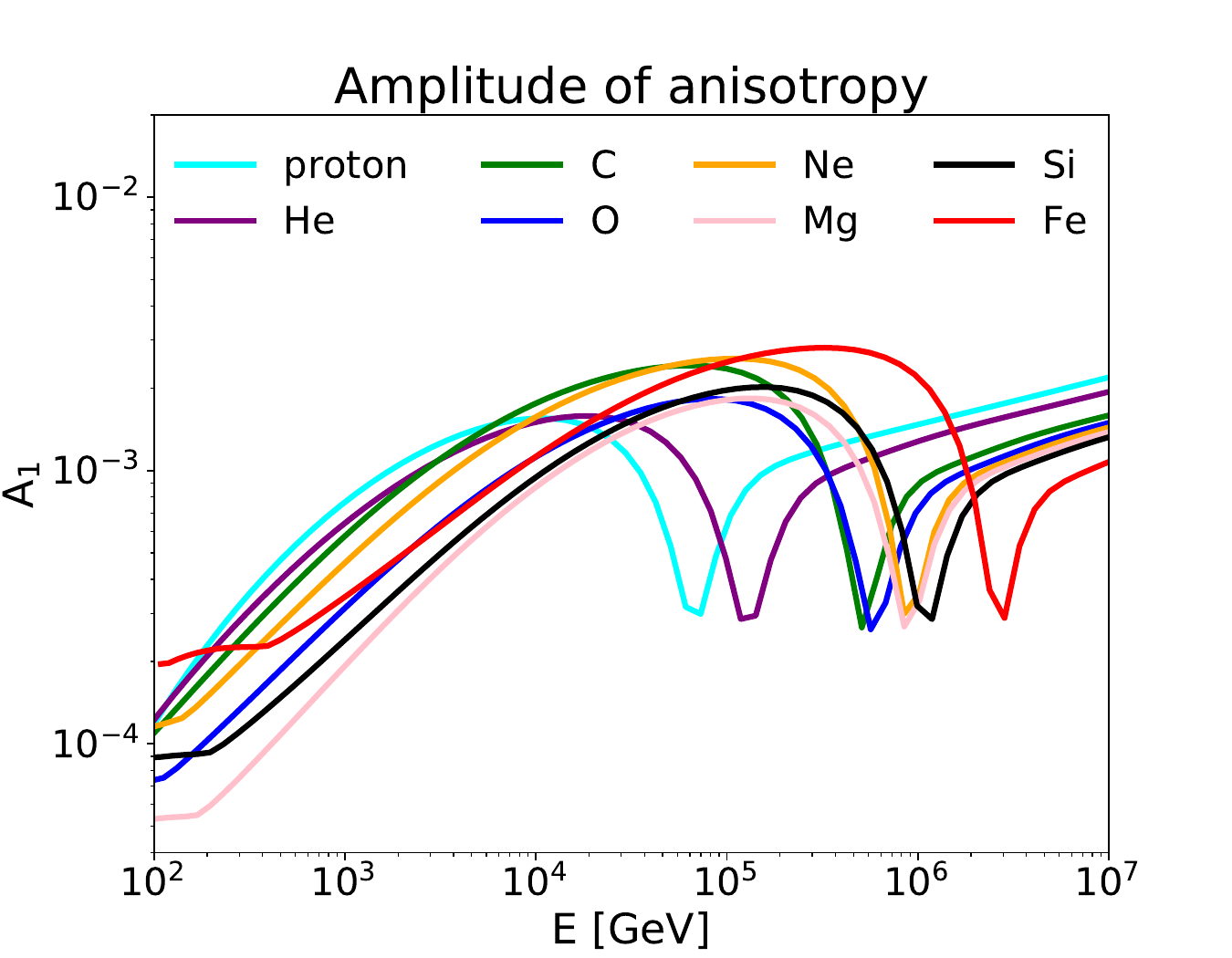}
        \label{fig:inj_ams}
    \end{subfigure}
    \hfill
    \begin{subfigure}[]{0.48\textwidth}
        \centering
        \includegraphics[
            width=\linewidth,
            height=7cm,                  
        ]{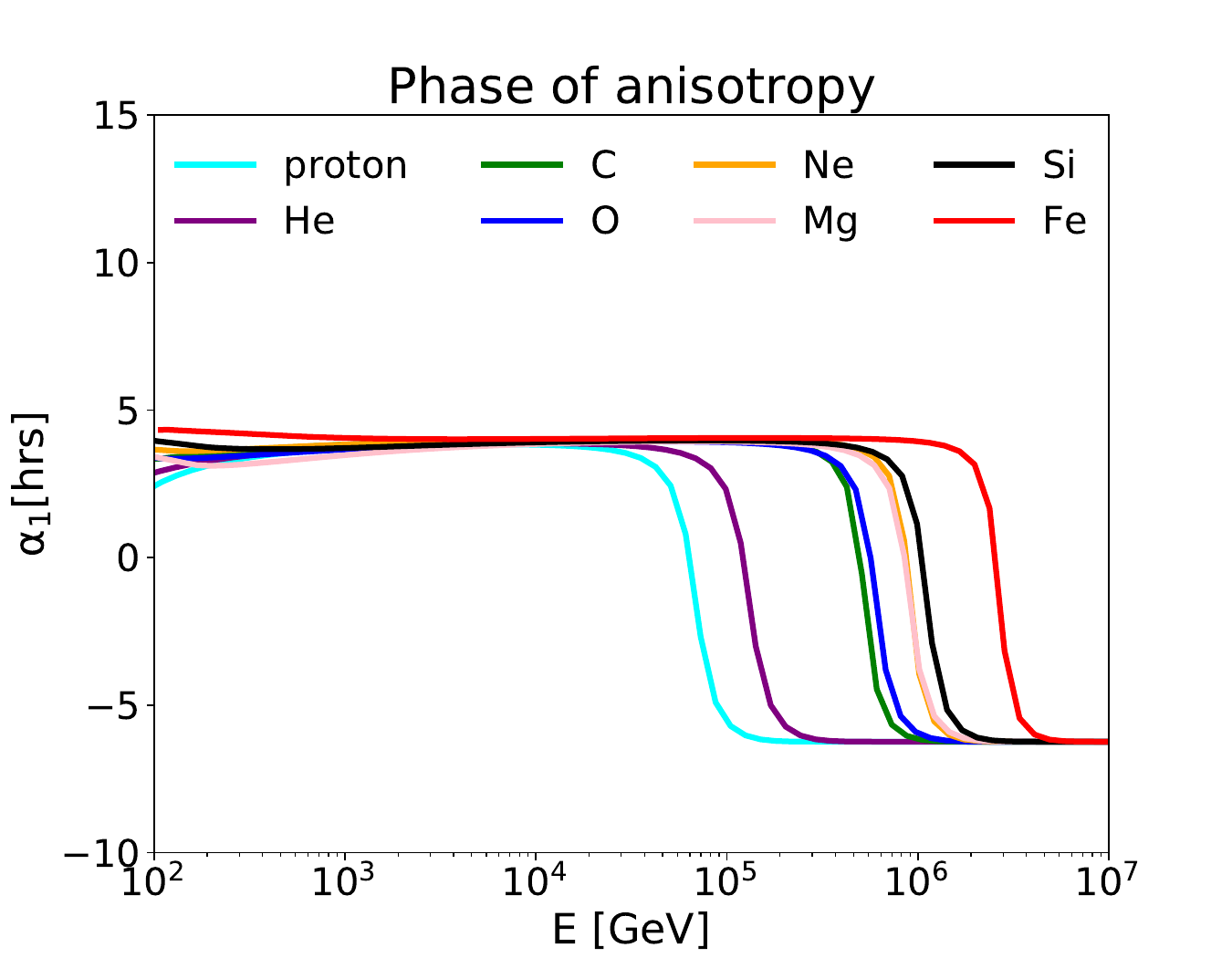}
        \label{fig:inj_dampe}
    \end{subfigure}
    \caption{Model-predicted amplitudes (left) and phases (right) of the anisotropies for individual elements.
     }
    \label{fig:injection_spectra}
\end{figure*}


\subsection{Spatial Diffusion}
Charged CR particles are scattered by turbulent magnetic fields during propagation, and their transport is generally described as a diffusion process. The diffusion coefficient $D_{xx}$ characterizes the efficiency of this spatial transport. In conventional CR propagation models, $D_{xx}$ is usually assumed to be spatially homogeneous and to depend only on particle rigidity. However, observations of extended emission around CR sources by HAWC \citep{abeysekara2017extended} and LHAASO \citep{aharonian2021extended} suggest that diffusion in source environments can be much slower than that in the average interstellar medium (ISM). This motivates the use of spatially dependent propagation (SDP) models, in which the diffusion coefficient varies with position and is correlated with the Galactic source distribution.

The SDP model adopted in this work follows the framework developed and applied in previous studies to describe Galactic CR spectral anomalies, large-scale anisotropy, and diffuse gamma-ray emission; see \citep{yao2024common,2024arXiv241209016D,PhysRevD.109.063027,2026arXiv260207610N,Nie:2024txr,2026ApJ...996...77Q} for more details. In this framework, the diffusion region is divided into inner and outer zones. The diffusion coefficient in the inner zone is anticorrelated with the source distribution, while that in the outer zone follows the conventional spatially independent form. Mathematically, the diffusion coefficient is parameterized as
\begin{equation}
D_{xx}(r,z,R)=D_{0}F(r,z)\beta^{\eta}
\left(\frac{R}{R_{0}}\right)^{\delta_0F(r,z)},
\end{equation}
where $D_0$ is the normalization at the reference rigidity $R_0=\SI{4}~$GV, $\beta=v/c$, and $\delta_0$ controls the rigidity dependence of diffusion. The factor $\beta^\eta$ accounts for low-energy corrections; in this work, we take $\beta=1$. The spatial modulation function $F(r,z)$ is defined as
\begin{equation}
F(r,z)=
\begin{cases}
g(r,z)+[1-g(r,z)]
\left(\frac{|z|}{\xi z_0}\right)^n,
& |z|\le \xi z_0, \\
1,
& |z|>\xi z_0 ,
\end{cases}
\end{equation}
with
\begin{equation}
g(r,z)=\frac{N_m}{1+f(r,z)} .
\end{equation}
Here, $N_m$ controls the normalization of the diffusion suppression, $\xi$ gives the fractional thickness of the inner halo, and $n$ controls the smoothness of the transition between the inner and outer zones. In this work, the spatial modulation of the diffusion coefficient is assumed to follow the large-scale Galactic source distribution, which we approximate using $f(r,z)=f_A(r,z)$.

We solve the transport equation using the numerical package DRAGON \citep{evoli2008cosmic}. The interstellar gas distribution, CO distribution, and secondary fluxes are calculated using the GALPROP reaction cross sections embedded in DRAGON \citep{evoli2008cosmic}. Solar modulation is included using the force-field approximation \citep{gleeson1968solar}. The key SDP parameters adopted in this work are summarized in Table~II.

\subsection{Mean Logarithmic Mass $\langle \ln A\rangle$}
The mass composition provides an important bridge between the component-resolved CR spectra obtained from direct space-borne measurements and the composition inferred from indirect ground-based air-shower observations. Given the spectra of individual species, the mean logarithmic mass at energy $E$ is calculated as
\begin{equation}
\langle \ln A \rangle(E)
=
\frac{
\sum_i \Phi_i(E)\ln A_i
}{
\sum_i \Phi_i(E)
},
\end{equation}
where $\Phi_i(E)$ is the flux of the $i$th nuclear species and $A_i$ is its atomic mass number. This quantity represents the flux-weighted average of $\ln A$ over all included nuclear species: lighter components such as protons give smaller values, while heavier components such as iron increase $\langle \ln A \rangle$.

In this work, we include the dominant primary nuclei, namely p, He, C, O, Ne, Mg, Si, and Fe. For the experimental estimate, the published component-resolved spectra from AMS-02 and DAMPE are interpolated onto a common energy grid and combined using the equation above. For the model prediction, the same expression is applied to the calculated fluxes of all nuclear species, including contributions from Type-A sources, Type-B sources, and the nearby source. The resulting $\langle \ln A \rangle$ distribution is then compared with the LHAASO composition measurements and used as an additional constraint on the model.

\subsection{Anisotropy}
In the diffusion picture, large-scale CR anisotropy originates from the nonuniform spatial distribution of CRs around the Solar System. If the local CR density were perfectly isotropic and homogeneous, particles would arrive at Earth uniformly from all directions. However, the Galactic source distribution, spatially dependent diffusion, and the contribution from a nearby source can generate a nonzero CR density gradient near the Solar position. This gradient drives a net diffusive streaming of CRs, which appears observationally as a dipole anisotropy.

For each source component and nuclear species, the propagated CR density
$\psi(R,\vec{r})$ is obtained from the transport calculation. The corresponding
dipole anisotropy is estimated from the local density gradient as
\begin{equation}
\boldsymbol{\Delta}(R)=
\frac{3D_{xx}(R,\vec{r}_{\odot})}{v\psi(R,\vec{r}_{\odot})}
\nabla \psi(R,\vec{r})\bigg|_{\vec{r}=\vec{r}_{\odot}},
\end{equation}
where $v$ is the particle velocity, $\psi(R,\vec{r}_{\odot})$ is the CR density at the Solar position, and $\nabla\psi$ is the local spatial gradient. This expression shows that the anisotropy amplitude is controlled by both the diffusion coefficient and the relative density gradient. A larger diffusion coefficient allows CRs to stream more efficiently, while a steeper local density gradient produces a stronger directional imbalance.

The total dipole anisotropy is calculated as a flux-weighted vector sum over all relevant source components and nuclear species,
\begin{equation}
\boldsymbol{\Delta}_{\rm tot}(E)
=\frac{
\sum_i \Phi_i(E)\boldsymbol{\Delta}_i(E)
}{
\sum_i \Phi_i(E)
},
\end{equation}
where $\Phi_i(E)$ and $\boldsymbol{\Delta}_i(E)$ are the flux and dipole anisotropy vector of the $i$th component, respectively. The vector nature of this summation is important because different components may point in different directions, leading to either enhancement or cancellation of the total anisotropy.

The anisotropy amplitude is then given by
\begin{equation}
A_1(E)=|\boldsymbol{\Delta}_{\rm tot}(E)|,
\end{equation}
and the phase is determined by the direction of $\boldsymbol{\Delta}_{\rm tot}(E)$. For comparison with observations, the calculated dipole direction is converted into right ascension and expressed in local sidereal time. The same procedure is applied to both the all-particle anisotropy and the component-resolved anisotropies of individual nuclear species.

\section{RESULTS and Discussion}
Based on the model setup described above, we first examine whether the adopted three-component source model can reproduce the component-resolved CR energy spectra. This step is essential because both the mean logarithmic mass and the anisotropy depend on the relative contributions of different nuclear species and source populations.
\subsection{Spectra for Individual Species}
Figure~1 presents the model prediction of the energy spectra of the major CR nuclei, including p, He, C, O, Ne, Mg, Si, and Fe, compared with measurements from AMS-02, DAMPE, and LHAASO. Overall, the model reproduces the main spectral features observed in the individual components over a broad energy range. The Type-A component dominates the low-energy background, the nearby source produces a broad excess at tens of TeV per nucleon and accounts for the observed spectral softening, while the harder Type-B component becomes important at higher energies and extends the spectra toward the sub-PeV to PeV region.

The DAMPE measurements play a particularly important role by bridging the energy range between AMS-02 and LHAASO. They constrain the cutoff of the nearby source and show that the spectral structures of different nuclei cannot be described by a single smooth power law. Instead, the observed spectra require the interplay of a low-energy background, a nearby-source contribution, and a high-energy Galactic component. Although the high-energy behavior of heavier nuclei is still limited by current statistics, the same three-component framework provides a consistent description of the available component-resolved spectra. Future LHAASO measurements of heavier nuclei in the sub-PeV to PeV range will be essential for validating this scenario and refining the injection parameters of the Type-B component.

\subsection{Mean Logarithmic Mass}

We next examine the energy evolution of the CR mass composition. The red points in Figure~2 show the $\langle \ln A\rangle$ values calculated from the component-resolved AMS-02 and DAMPE spectra, while the blue points show the LHAASO composition measurements. The two data sets connect smoothly in the overlapping energy region, providing a continuous description of the composition evolution from direct measurements to the air-shower regime. The model reproduces the overall trend well, including the nonmonotonic structure around $\sim 100$~TeV.

The component decomposition shows that the low-energy composition is mainly shaped by the Type-A component, while the bump-like feature around $\sim 100$~TeV is largely driven by the nearby-source contribution. At higher energies, the nearby-source contribution gradually cuts off, and the Type-B component becomes increasingly important toward the LHAASO energy range. Therefore, the evolution of $\langle \ln A\rangle$ provides an important consistency check of the three-component source picture: the same components that reproduce the individual spectra can also account for the observed mass-composition evolution.

\subsection{Anisotropy}

We further examine whether the same source components can account for the observed large-scale CR anisotropy. The steady-state Type-A and Type-B populations mainly trace the large-scale Galactic CR gradient shaped by the global source distribution and propagation environment. In contrast, the nearby source is localized and time dependent, and can therefore generate a sizable local gradient even when its flux contribution is not dominant. As a result, the total anisotropy is sensitive to the competition between the large-scale Galactic background and the nearby-source contribution.

Figure~3 compares the model-calculated all-particle anisotropy with observational data. The left panel shows the dipole amplitude, while the right panel shows the phase. The Type-A and Type-B background components provide the large-scale Galactic contribution, whereas the nearby source introduces an additional anisotropy component whose direction is determined by its relative position to the Solar System. The vector combination of these components can lead to either enhancement or partial cancellation of the total dipole anisotropy. Overall, the model reproduces the observed energy-dependent behavior of both the amplitude and phase, indicating that the nearby source plays an important role in shaping the anisotropy features.

Using the component-resolved spectra constrained by AMS-02, DAMPE, and LHAASO, we also predict the anisotropy of individual nuclear species. Figure~4 shows the model-predicted dipole amplitudes and phases for p, He, C, O, Ne, Mg, Si, and Fe. The transition features appear at different total energies for different nuclei, reflecting the rigidity-dependent nature of the source spectra and propagation. In particular, heavier nuclei exhibit similar anisotropy behavior shifted to higher total energies according to their charge number (Z). Future component-resolved anisotropy measurements, especially from LHAASO, will provide a critical test of this prediction and of the proposed common origin of the spectral, composition, and anisotropy features.

\section{Summary}

In this work, we investigate whether the observed spectral features, mass-composition evolution, and large-scale anisotropy of Galactic CRs can be understood within a unified physical framework. Motivated by the latest DAMPE component-resolved spectra from protons to iron and LHAASO measurements of mass composition in the sub-PeV to PeV region, we adopted a spatially dependent propagation model with three source populations: a conventional Type-A Galactic background, a higher-energy Type-B component, and a nearby-source contribution.

We demonstrated that this model can reproduce the main spectral features of individual CR nuclei, including p, He, C, O, Ne, Mg, Si, and Fe, over a broad energy range. The Type-A component dominates the low-energy spectra, the nearby source accounts for the broad excess and spectral softening around tens of TeV per nucleon, and the Type-B component becomes important at higher energies, extending the spectra toward the sub-PeV to PeV range. The DAMPE measurements play a key role in bridging the energy range between AMS-02 and LHAASO and provide important constraints on the cutoff of the nearby-source contribution.

Using the component-resolved spectra from AMS-02 and DAMPE, we calculated the mean logarithmic mass, $\langle \ln A\rangle$, and found that it connects smoothly to the LHAASO composition measurements. The resulting $\langle \ln A\rangle$ distribution exhibits a nonmonotonic structure around $\sim 100$~TeV, which is reproduced by the model. The component decomposition indicates that this feature is mainly driven by the nearby source, while the high-energy rise toward the LHAASO range is associated with the increasing contribution from the Type-B component.

We further calculated the large-scale dipole anisotropy using the same source and propagation setup. The observed energy-dependent behavior of the anisotropy amplitude and phase can be described by the competition between the large-scale Galactic background and the nearby-source contribution. We also predicted the component-resolved anisotropy for individual nuclear species. The transition features in the amplitude and phase shift to higher total energies for heavier nuclei, reflecting the rigidity-dependent nature of the source spectra and propagation.

These results suggest that the observed CR spectral features, mass-composition evolution, and anisotropy can be consistently interpreted within a common physical scenario. In this scenario, the nearby source plays a central role in shaping the spectral softening, the $\langle \ln A\rangle$ bump, and the anisotropy behavior from the TeV to sub-PeV region, while the Type-B component controls the high-energy continuation. We note, however, that the present model is phenomenological, and the decomposition into Type-A, Type-B, and nearby-source components may not be unique. The interpretation also depends on systematic uncertainties in composition reconstruction. Future LHAASO measurements of heavier nuclei and component-resolved anisotropy will therefore provide crucial tests of this unified picture and further constrain the injection properties of the high-energy Galactic component.

\begin{acknowledgments}
X. Dong performed the model calculations and prepared the initial draft. Y. Yao contributed to the scientific motivation, interpretation of the results, and manuscript writing and revision. Y. Guo proposed the main idea of the study, developed the theoretical framework, guided the interpretation of the propagation model, and supervised the project. S. Cui contributed to scientific discussions and project supervision.

This work was financially supported by the National Key R\&D Program of China (Grant No. 2025SKA0110103), the National Natural Science Foundation of China (Grants No. 12333006, 12275279, 12373105, and 12320101005), and the Hebei Province Graduate Student Innovation Project (Grant No. CXZZBS2026083).
 \end{acknowledgments}

\bibliographystyle{unsrt_update}
\bibliography{apssamp}

\providecommand{\noopsort}[1]{}\providecommand{\singleletter}[1]{#1}%
\begin{thebibliography}{10}

\bibitem{2020PhR...872....1B}
Julia {Becker Tjus} and Lukas {Merten}.
\newblock {Closing in on the origin of Galactic cosmic rays using
  multimessenger information}.
\newblock {\em \physrep}, 872:1--98, August 2020.

\bibitem{1983RPPh...46..973D}
L.~Oc. {Drury}.
\newblock {REVIEW ARTICLE: An introduction to the theory of diffusive shock
  acceleration of energetic particles in tenuous plasmas}.
\newblock {\em Reports on Progress in Physics}, 46(8):973--1027, August 1983.

\bibitem{2007ARNPS..57..285S}
Andrew~W. {Strong}, Igor~V. {Moskalenko}, and Vladimir~S. {Ptuskin}.
\newblock {Cosmic-Ray Propagation and Interactions in the Galaxy}.
\newblock {\em Annual Review of Nuclear and Particle Science}, 57(1):285--327,
  November 2007.

\bibitem{2009BRASP..73..564P}
A.~D. {Panov}, J.~H. {Adams}, H.~S. {Ahn}, et~al.
\newblock {Energy spectra of abundant nuclei of primary cosmic rays from the
  data of ATIC-2 experiment: Final results}.
\newblock {\em Bulletin of the Russian Academy of Sciences, Physics},
  73(5):564--567, June 2009.

\bibitem{2011Sci...332...69A}
O.~{Adriani}, G.~C. {Barbarino}, G.~A. {Bazilevskaya}, et~al.
\newblock {PAMELA Measurements of Cosmic-Ray Proton and Helium Spectra}.
\newblock {\em Science}, 332(6025):69, April 2011.

\bibitem{2011ApJ...728..122Y}
Y.~S. {Yoon}, H.~S. {Ahn}, P.~S. {Allison}, et~al.
\newblock {Cosmic-ray Proton and Helium Spectra from the First CREAM Flight}.
\newblock {\em \apj}, 728(2):122, February 2011.

\bibitem{2017PhRvL.119y1101A}
M.~{Aguilar}, L.~{Ali Cavasonza}, B.~{Alpat}, et~al.
\newblock {Observation of the Identical Rigidity Dependence of He, C, and O
  Cosmic Rays at High Rigidities by the Alpha Magnetic Spectrometer on the
  International Space Station}.
\newblock {\em \prl}, 119(25):251101, December 2017.

\bibitem{2018PhRvL.120b1101A}
M.~{Aguilar}, L.~{Ali Cavasonza}, G.~{Ambrosi}, et~al.
\newblock {Observation of New Properties of Secondary Cosmic Rays Lithium,
  Beryllium, and Boron by the Alpha Magnetic Spectrometer on the International
  Space Station}.
\newblock {\em \prl}, 120(2):021101, January 2018.

\bibitem{2019SciA....5.3793A}
Q.~{An}, R.~{Asfandiyarov}, P.~{Azzarello}, et~al.
\newblock {Measurement of the cosmic ray proton spectrum from 40 GeV to 100 TeV
  with the DAMPE satellite}.
\newblock {\em Science Advances}, 5(9):eaax3793, September 2019.

\bibitem{2021PhRvL.126t1102A}
F.~{Alemanno}, Q.~{An}, P.~{Azzarello}, et~al.
\newblock {Measurement of the Cosmic Ray Helium Energy Spectrum from 70 GeV to
  80 TeV with the DAMPE Space Mission}.
\newblock {\em \prl}, 126(20):201102, May 2021.

\bibitem{2019PhRvL.122r1102A}
O.~{Adriani}, Y.~{Akaike}, K.~{Asano}, et~al.
\newblock {Direct Measurement of the Cosmic-Ray Proton Spectrum from 50 GeV to
  10 TeV with the Calorimetric Electron Telescope on the International Space
  Station}.
\newblock {\em \prl}, 122(18):181102, May 2019.

\bibitem{2020PhRvL.125y1102A}
O.~{Adriani}, Y.~{Akaike}, K.~{Asano}, et~al.
\newblock {Direct Measurement of the Cosmic-Ray Carbon and Oxygen Spectra from
  10 GeV /n to 2.2 TeV /n with the Calorimetric Electron Telescope on the
  International Space Station}.
\newblock {\em \prl}, 125(25):251102, December 2020.

\bibitem{2022PhRvL.129j1102A}
O.~{Adriani}, Y.~{Akaike}, K.~{Asano}, et~al.
\newblock {Observation of Spectral Structures in the Flux of Cosmic-Ray Protons
  from 50 GeV to 60 TeV with the Calorimetric Electron Telescope on the
  International Space Station}.
\newblock {\em \prl}, 129(10):101102, September 2022.

\bibitem{2023PhRvL.130q1002A}
O.~{Adriani}, Y.~{Akaike}, K.~{Asano}, et~al.
\newblock {Direct Measurement of the Cosmic-Ray Helium Spectrum from 40 GeV to
  250 TeV with the Calorimetric Electron Telescope on the International Space
  Station}.
\newblock {\em \prl}, 130(17):171002, April 2023.

\bibitem{2011PhRvD..84d3002Y}
Qiang {Yuan}, Bing {Zhang}, and Xiao-Jun {Bi}.
\newblock {Cosmic ray spectral hardening due to dispersion in the source
  injection spectra}.
\newblock {\em \prd}, 84(4):043002, August 2011.

\bibitem{2019SCPMA..6249511Y}
Qiang {Yuan}.
\newblock {Implications on cosmic ray injection and propagation parameters from
  Voyager/ACE/AMS-02 nucleus data}.
\newblock {\em Science China Physics, Mechanics, and Astronomy}, 62(4):49511,
  April 2019.

\bibitem{2012MNRAS.421.1209T}
Satyendra {Thoudam} and J{\"o}rg~R. {H{\"o}randel}.
\newblock {Nearby supernova remnants and the cosmic ray spectral hardening at
  high energies}.
\newblock {\em \mnras}, 421(2):1209--1214, April 2012.

\bibitem{pzxy-v9v8}
Qiang Yuan.
\newblock Implication of multiple source populations of galactic cosmic rays
  from proton and helium spectra.
\newblock {\em Phys. Rev. D}, May 2026.

\bibitem{Yue_2019}
Chuan Yue, Peng-Xiong Ma, Qiang Yuan, et~al.
\newblock Implications on the origin of cosmic rays in light of 10 tv spectral
  softenings.
\newblock {\em Frontiers of Physics}, 15(2), December 2019.

\bibitem{2012PhRvL.109f1101B}
Pasquale {Blasi}, Elena {Amato}, and Pasquale~D. {Serpico}.
\newblock {Spectral Breaks as a Signature of Cosmic Ray Induced Turbulence in
  the Galaxy}.
\newblock {\em \prl}, 109(6):061101, August 2012.

\bibitem{Yuan_2017}
Qiang Yuan, Su-Jie Lin, Kun Fang, and Xiao-Jun Bi.
\newblock Propagation of cosmic rays in the ams-02 era.
\newblock {\em Physical Review D}, 95(8), April 2017.

\bibitem{2012ApJ...752L..13T}
Nicola {Tomassetti}.
\newblock {Origin of the Cosmic-Ray Spectral Hardening}.
\newblock {\em \apjl}, 752(1):L13, June 2012.

\bibitem{2016ApJ...819...54G}
Yi-Qing {Guo}, Zhen {Tian}, and Chao {Jin}.
\newblock {Spatial-dependent Propagation of Cosmic Rays Results in the Spectrum
  of Proton, Ratios of P/P, and B/C, and Anisotropy of Nuclei}.
\newblock {\em \apj}, 819(1):54, March 2016.

\bibitem{2024PhRvD.109l1101A}
F.~{Alemanno}, C.~{Altomare}, Q.~{An}, et~al.
\newblock {Measurement of the cosmic p +He energy spectrum from 50 GeV to 0.5
  PeV with the DAMPE space mission}.
\newblock {\em \prd}, 109(12):L121101, June 2024.

\bibitem{2026Natur.653...52D}
{Dampe Collaboration}, Francesca {Alemanno}, Qi~{An}, et~al.
\newblock {Charge-dependent spectral softenings of primary cosmic rays below
  the knee}.
\newblock {\em \nat}, 653(8113):52--55, May 2026.

\bibitem{2024PhRvL.132e1002V}
F.~{Varsi}, S.~{Ahmad}, M.~{Chakraborty}, et~al.
\newblock {Evidence of a Hardening in the Cosmic Ray Proton Spectrum at around
  166 TeV Observed by the GRAPES-3 Experiment}.
\newblock {\em \prl}, 132(5):051002, January 2024.

\bibitem{2025SciBu..70.4173C}
Zhen {Cao}, F.~{Aharonian}, Y.~X. {Bai}, et~al.
\newblock {Precise measurements of the cosmic ray proton energy spectrum in the
  ``knee'' region}.
\newblock {\em Science Bulletin}, 70(24):4173--4180, December 2025.

\bibitem{2026PhRvL.136l1001C}
Zhen {Cao}, F.~{Aharonian}, Y.~X. {Bai}, et~al.
\newblock {Precise Measurement of the Cosmic Ray Helium Spectrum above 0.1
  PeV}.
\newblock {\em \prl}, 136(12):121001, March 2026.

\bibitem{2024PhRvL.132m1002C}
Zhen {Cao}, F.~{Aharonian}, {Axikegu}, et~al.
\newblock {Measurements of All-Particle Energy Spectrum and Mean Logarithmic
  Mass of Cosmic Rays from 0.3 to 30 PeV with LHAASO-KM2A}.
\newblock {\em \prl}, 132(13):131002, March 2024.

\bibitem{2016PhRvL.117o1103A}
Markus {Ahlers}.
\newblock {Deciphering the Dipole Anisotropy of Galactic Cosmic Rays}.
\newblock {\em \prl}, 117(15):151103, October 2016.

\bibitem{2012JCAP...01..011B}
Pasquale {Blasi} and Elena {Amato}.
\newblock {Diffusive propagation of cosmic rays from supernova remnants in the
  Galaxy. II: anisotropy}.
\newblock {\em \jcap}, 2012(1):011, January 2012.

\bibitem{2009ApJ...698.2121A}
A.~A. {Abdo}, B.~T. {Allen}, T.~{Aune}, et~al.
\newblock {The Large-Scale Cosmic-Ray Anisotropy as Observed with Milagro}.
\newblock {\em \apj}, 698(2):2121--2130, June 2009.

\bibitem{2012ApJ...746...33A}
R.~{Abbasi}, Y.~{Abdou}, T.~{Abu-Zayyad}, et~al.
\newblock {Observation of Anisotropy in the Galactic Cosmic-Ray Arrival
  Directions at 400 TeV with IceCube}.
\newblock {\em \apj}, 746(1):33, February 2012.

\bibitem{2015ApJ...809...90B}
B.~{Bartoli}, P.~{Bernardini}, X.~J. {Bi}, et~al.
\newblock {ARGO-YBJ Observation of the Large-scale Cosmic Ray Anisotropy During
  the Solar Minimum between Cycles 23 and 24}.
\newblock {\em \apj}, 809(1):90, August 2015.

\bibitem{2017ApJ...836..153A}
M.~{Amenomori}, X.~J. {Bi}, D.~{Chen}, et~al.
\newblock {Northern Sky Galactic Cosmic Ray Anisotropy between 10 and 1000 TeV
  with the Tibet Air Shower Array}.
\newblock {\em \apj}, 836(2):153, February 2017.

\bibitem{2017PrPNP..94..184A}
Markus {Ahlers} and Philipp {Mertsch}.
\newblock {Origin of small-scale anisotropies in Galactic cosmic rays}.
\newblock {\em Progress in Particle and Nuclear Physics}, 94:184--216, May
  2017.

\bibitem{LHAASO2025ProtonAnisotropy}
{LHAASO Collaboration}.
\newblock Observation of large-scale anisotropy of very high-energy cosmic-ray
  protons with lhaaso-km2a.
\newblock In {\em Proceedings of 39th International Cosmic Ray Conference ---
  ICRC2025}, volume 501, page 286, 2025.

\bibitem{2019FrASS...6...23B}
Rafael {Alves Batista}, Jonathan {Biteau}, Mauricio {Bustamante}, et~al.
\newblock {Open Questions in Cosmic-Ray Research at Ultrahigh Energies}.
\newblock {\em Frontiers in Astronomy and Space Sciences}, 6:23, June 2019.

\bibitem{2024JCAP...01..022A}
A.~{Abdul Halim}, P.~{Abreu}, M.~{Aglietta}, et~al.
\newblock {Constraining models for the origin of ultra-high-energy cosmic rays
  with a novel combined analysis of arrival directions, spectrum, and
  composition data measured at the Pierre Auger Observatory}.
\newblock {\em \jcap}, 2024(1):022, January 2024.

\bibitem{qiao2026coevolutioncosmicrayenergy}
Bing-Qiang Qiao, Qiang Yuan, and Yi-Qing Guo.
\newblock Co-evolution of cosmic ray energy spectra, composition, and
  anisotropies, 2026.

\bibitem{PhysRevD.97.063008}
Yi-Qing Guo and Qiang Yuan.
\newblock Understanding the spectral hardenings and radial distribution of
  galactic cosmic rays and fermi diffuse $\ensuremath{\gamma}$ rays with
  spatially-dependent propagation.
\newblock {\em Phys. Rev. D}, 97:063008, Mar 2018.

\bibitem{2023ApJ...956...75Q}
Bing-Qiang {Qiao}, Yi-Qing {Guo}, Wei {Liu}, and Xiao-Jun {Bi}.
\newblock {Nearby SNR: A Possible Common Origin of Multi-messenger Anomalies in
  the Spectra, Ratios, and Anisotropy of Cosmic Rays}.
\newblock {\em \apj}, 956(2):75, October 2023.

\bibitem{2026ApJ...996...77Q}
Bing-Qiang {Qiao}, Wei {Liu}, Huirong {Yan}, and Yi-Qing {Guo}.
\newblock {The Compton-Getting Origin of the Large-scale Anisotropy of Galactic
  Cosmic Rays}.
\newblock {\em \apj}, 996(1):77, January 2026.

\bibitem{maurin2002galactic}
David Maurin, Richard Taillet, Fiorenza Donato, et~al.
\newblock Galactic cosmic ray nuclei as a tool for astroparticle physics.
\newblock {\em arXiv preprint astro-ph/0212111}, 2002.

\bibitem{2025NSRev..12af496L}
{Lhaaso Collaboration}, Zhen {Cao}, Felix {Aharonian}, et~al.
\newblock {Ultrahigh-Energy Gamma-ray Emission Associated with Black Hole-Jet
  Systems}.
\newblock {\em National Science Review}, 12(12):nwaf496, December 2025.

\bibitem{2025arXiv251216638T}
{The LHAASO Collaboration}, Zhen {Cao}, F.~{Aharonian}, et~al.
\newblock {Cygnus X-3: A variable petaelectronvolt gamma-ray source}.
\newblock {\em arXiv e-prints}, page arXiv:2512.16638, December 2025.

\bibitem{1996A&AS..120C.437C}
G.~{Case} and D.~{Bhattacharya}.
\newblock {Revisiting the galactic supernova remnant distribution.}
\newblock {\em aaps}, 120:437--440, December 1996.

\bibitem{2026ApJ...997..163Y}
Hua {Yue}, Jianli {Zhang}, Yuhai {Ge}, et~al.
\newblock {X-Ray Binaries: A Potential Dominant Contributor to the Cosmic-Ray
  Spectral Knee Structure}.
\newblock {\em \apj}, 997(2):163, February 2026.

\bibitem{aguilar2021alpha}
M~Aguilar, L~Ali Cavasonza, G~Ambrosi, et~al.
\newblock The alpha magnetic spectrometer (ams) on the international space
  station: Part ii—results from the first seven years.
\newblock {\em Physics reports}, 894:1--116, 2021.

\bibitem{aguilar2024properties}
M~Aguilar, B~Alpat, G~Ambrosi, et~al.
\newblock Properties of cosmic deuterons measured by the alpha magnetic
  spectrometer.
\newblock {\em Physical review letters}, 132(26):261001, 2024.

\bibitem{aguilar2023properties}
M~Aguilar, L~Ali~Cavasonza, B~Alpat, et~al.
\newblock Properties of cosmic-ray sulfur and determination of the composition
  of primary cosmic-ray carbon, neon, magnesium, and sulfur: Ten-year results
  from the alpha magnetic spectrometer.
\newblock {\em Physical review letters}, 130(21):211002, 2023.

\bibitem{aguilar2021properties1}
M~Aguilar, L~Ali Cavasonza, MS~Allen, et~al.
\newblock Properties of iron primary cosmic rays: Results from the alpha
  magnetic spectrometer.
\newblock {\em Physical review letters}, 126(4):041104, 2021.

\bibitem{aguilar2020properties}
M~Aguilar, L~Ali~Cavasonza, G~Ambrosi, et~al.
\newblock Properties of neon, magnesium, and silicon primary cosmic rays
  results from the alpha magnetic spectrometer.
\newblock {\em Physical review letters}, 124(21):211102, 2020.

\bibitem{aguilar2019properties}
M~Aguilar, L~Ali~Cavasonza, G~Ambrosi, et~al.
\newblock Properties of cosmic helium isotopes measured by the alpha magnetic
  spectrometer.
\newblock {\em Physical review letters}, 123(18):181102, 2019.

\bibitem{Casilli:2025w8}
Elisabetta Casilli, Francesca Alemanno, Paolo Bernardini, et~al.
\newblock {Direct measurement of cosmic neon, magnesium, and silicon fluxes
  with DAMPE}.
\newblock {\em PoS}, ICRC2025:024, 2025.

\bibitem{1973ICRC....2.1058S}
S.~{Sakakibara}, H.~{Ueno}, K.~{Fujimoto}, I.~{Kondo}, and K.~{Nagashima}.
\newblock {Sidereal Time Variation of Small Air Showers Observed at Mt.
  Norikura.}
\newblock In {\em International Cosmic Ray Conference}, volume~2 of {\em
  International Cosmic Ray Conference}, page 1058, January 1973.

\bibitem{1981ICRC...10..246B}
M.~{Bercovitch} and S.~P. {Agrawal}.
\newblock {Cosmic ray anisotropies at median primary rigidities between 100 and
  1000 GV}.
\newblock In {\em International Cosmic Ray Conference}, volume~10 of {\em
  International Cosmic Ray Conference}, pages 246--249, January 1981.

\bibitem{1983ICRC....3..383T}
T.~{Thambyahpillai}.
\newblock {The Sidereal Diurnal Variation Measured Underground in London}.
\newblock In {\em International Cosmic Ray Conference}, volume~3 of {\em
  International Cosmic Ray Conference}, page 383, August 1983.

\bibitem{1985P&SS...33.1069S}
D.~B. {Swinson} and K.~{Nagashima}.
\newblock {Corrected sidereal anisotropy for underground muons}.
\newblock {\em \planss}, 33(9):1069--1072, September 1985.

\bibitem{1987ICRC....2...22A}
Yu.~M. {Andreyev}, A.~E. {Chudakov}, V.~A. {Kozyarivsky}, et~al.
\newblock {Cosmic Ray Sidereal Anisotropy Observed by Baksan Underground Muon
  Telescope}.
\newblock In {\em International Cosmic Ray Conference}, volume~2 of {\em
  International Cosmic Ray Conference}, page~22, January 1987.

\bibitem{1987ICRC....2...18L}
Y.~W. {Lee} and L.~K. {Ng}.
\newblock {Observation of Cosmic-Ray Intensity Variation Using AN Underground
  Telescope}.
\newblock In {\em International Cosmic Ray Conference}, volume~2 of {\em
  International Cosmic Ray Conference}, page~18, January 1987.

\bibitem{1990ICRC....6..361U}
H.~{Ueno}, Z.~{Fujii}, and T.~{Yamada}.
\newblock {11 Years Variations of Sidereal Anisotropy Observed at Sakashita
  Underground Station}.
\newblock In {\em 21st International Cosmic Ray Conference (ICRC21)}, volume~6
  of {\em International Cosmic Ray Conference}, page 361, January 1990.

\bibitem{1991ApJ...376..322C}
D.~J. {Cutler} and D.~E. {Groom}.
\newblock {Mayflower Mine 1500 GV Detector: Cosmic-Ray Anisotropy and Search
  for Cygnus X-3}.
\newblock {\em \apj}, 376:322, July 1991.

\bibitem{1995ICRC....4..648M}
S.~{Mori}, S.~{Yasue}, K.~{Munakata}, et~al.
\newblock {Observation of Sidereal Anisotropy of Cosmic Rays at
  {\ensuremath{\sim}}1 TV}.
\newblock In {\em International Cosmic Ray Conference}, volume~4 of {\em
  International Cosmic Ray Conference}, page 648, January 1995.

\bibitem{1995ICRC....4..639M}
K.~{Munakata}, S.~{Yasue}, S.~{Mori}, et~al.
\newblock {Two Hemisphere Observations of the North-South Sidereal Asymmetry at
  {\ensuremath{\sim}}1 TeV}.
\newblock In {\em International Cosmic Ray Conference}, volume~4 of {\em
  International Cosmic Ray Conference}, page 639, January 1995.

\bibitem{1995ICRC....4..635F}
K.~B. {Fenton}, A.~G. {Fenton}, and J.~E. {Humble}.
\newblock {Sidereal Variations at High Energies - Observations at Poatina}.
\newblock In {\em International Cosmic Ray Conference}, volume~4 of {\em
  International Cosmic Ray Conference}, page 635, January 1995.

\bibitem{1997PhRvD..56...23M}
K.~{Munakata}, T.~{Kiuchi}, S.~{Yasue}, et~al.
\newblock {Large-scale anisotropy of the cosmic-ray muon flux in Kamiokande}.
\newblock {\em \prd}, 56(1):23--26, July 1997.

\bibitem{2003PhRvD..67d2002A}
M.~{Ambrosio}, R.~{Antolini}, A.~{Baldini}, et~al.
\newblock {Search for the sidereal and solar diurnal modulations in the total
  MACRO muon data set}.
\newblock {\em \prd}, 67(4):042002, February 2003.

\bibitem{2007PhRvD..75f2003G}
G.~{Guillian}, J.~{Hosaka}, K.~{Ishihara}, et~al.
\newblock {Observation of the anisotropy of 10TeV primary cosmic ray nuclei
  flux with the Super-Kamiokande-I detector}.
\newblock {\em \prd}, 75(6):062003, March 2007.

\bibitem{1975ICRC....2..586G}
T.~{Gombosi}, J.~{K{\'o}ta}, A.~J. {Somogyi}, et~al.
\newblock {Galactic cosmic ray anisotropy at
  {\ensuremath{\approx}}6{\texttimes}{}10$^{13}$eV.}
\newblock In {\em International Cosmic Ray Conference}, volume~2 of {\em
  International Cosmic Ray Conference}, pages 586--591, August 1975.

\bibitem{1981ICRC....2..146A}
V.~V. {Alexeyenko}, A.~E. {Chudakov}, E.~N. {Gulieva}, and V.~G. {Sborschikov}.
\newblock {Anisotropy of Small EAS (about 10(13) Ev)}.
\newblock In {\em International Cosmic Ray Conference}, volume~2 of {\em
  International Cosmic Ray Conference}, page 146, January 1981.

\bibitem{1989NCimC..12..695N}
K.~{Nagashima}, K.~{Fujimoto}, S.~{Sakakibara}, et~al.
\newblock {Galactic cosmic-ray anisotropy and its modulation in the
  heliomagnetosphere, inferred from air shower observation at Mt. Norikura.}
\newblock {\em Nuovo Cimento C Geophysics Space Physics C}, 12:695--749,
  December 1989.

\bibitem{1995ICRC....2..800A}
M.~{Aglietta}, B.~{Alessandro}, P.~{Antonioli}, et~al.
\newblock {Study of the Cosmic Ray Anisotropy at Eo {\ensuremath{\sim}} 100 TeV
  from EAS-TOP: 1992-1994}.
\newblock In {\em International Cosmic Ray Conference}, volume~2 of {\em
  International Cosmic Ray Conference}, page 800, January 1995.

\bibitem{1996ApJ...470..501A}
M.~{Aglietta}, B.~{Alessandro}, P.~{Antonioli}, et~al.
\newblock {A Measurement of the Solar and Sidereal Cosmic-Ray Anisotropy at E 0
  approximately 10 14 eV}.
\newblock {\em \apj}, 470:501, October 1996.

\bibitem{2009ApJ...692L.130A}
M.~{Aglietta}, V.~V. {Alekseenko}, B.~{Alessandro}, et~al.
\newblock {Evolution of the Cosmic-Ray Anisotropy Above {}10$^{14}$ eV}.
\newblock {\em \apjl}, 692(2):L130--L133, February 2009.

\bibitem{alekseenko200910}
VV~Alekseenko, AB~Cherniaev, DD~Djappuev, et~al.
\newblock 10-100 tev cosmic ray anisotropy measured at the baksan eas
  “carpet” array.
\newblock {\em Nuclear Physics B-Proceedings Supplements}, 196:179--182, 2009.

\bibitem{2015ICRC...34..281C}
A.~{Chiavassa}, W.~D. {Apel}, J.~C. {Arteaga-Vel{\'a}zquez}, et~al.
\newblock {A study of the first harmonic of the large scale anisotropies with
  the KASCADE-Grande experiment}.
\newblock In {\em 34th International Cosmic Ray Conference (ICRC2015)},
  volume~34 of {\em International Cosmic Ray Conference}, page 281, July 2015.

\bibitem{2010ApJ...718L.194A}
R.~{Abbasi}, Y.~{Abdou}, T.~{Abu-Zayyad}, et~al.
\newblock {Measurement of the Anisotropy of Cosmic-ray Arrival Directions with
  IceCube}.
\newblock {\em \apjl}, 718(2):L194--L198, August 2010.

\bibitem{2013ApJ...765...55A}
M.~G. {Aartsen}, R.~{Abbasi}, Y.~{Abdou}, et~al.
\newblock {Observation of Cosmic-Ray Anisotropy with the IceTop Air Shower
  Array}.
\newblock {\em \apj}, 765(1):55, March 2013.

\bibitem{2005ApJ...626L..29A}
M.~{Amenomori}, S.~{Ayabe}, S.~W. {Cui}, et~al.
\newblock {Large-Scale Sidereal Anisotropy of Galactic Cosmic-Ray Intensity
  Observed by the Tibet Air Shower Array}.
\newblock {\em \apjl}, 626(1):L29--L32, June 2005.

\bibitem{2015ICRC...34..355A}
M.~{Amenomori}, X.~J. {Bi}, D.~{Chen}, et~al.
\newblock {Northern sky Galactic Cosmic Ray anisotropy between 10-1000 TeV with
  the Tibet Air Shower Array}.
\newblock In {\em 34th International Cosmic Ray Conference (ICRC2015)},
  volume~34 of {\em International Cosmic Ray Conference}, page 355, July 2015.

\bibitem{2020ApJ...891..142A}
A.~{Aab}, P.~{Abreu}, M.~{Aglietta}, et~al.
\newblock {Cosmic-Ray Anisotropies in Right Ascension Measured by the Pierre
  Auger Observatory}.
\newblock {\em \apj}, 891(2):142, March 2020.

\bibitem{abeysekara2017extended}
AU~Abeysekara, A~Albert, R~Alfaro, et~al.
\newblock Extended gamma-ray sources around pulsars constrain the origin of the
  positron flux at earth.
\newblock {\em Science}, 358(6365):911--914, 2017.

\bibitem{aharonian2021extended}
F~Aharonian, Q~An, Axikegu, et~al.
\newblock Extended very-high-energy gamma-ray emission surrounding psr j 0622+
  3749 observed by lhaaso-km2a.
\newblock {\em Physical Review Letters}, 126(24):241103, 2021.

\bibitem{yao2024common}
Yu-Hua Yao, Xu-Lin Dong, Yi-Qing Guo, and Qiang Yuan.
\newblock Common origin of the multimessenger spectral anomaly of galactic
  cosmic rays.
\newblock {\em Physical Review D}, 109(6):063001, 2024.

\bibitem{2024arXiv241209016D}
Xu-Lin {Dong}, Shu-Wei {Ma}, Yi-Qing {Guo}, and Shu-Wang {Cui}.
\newblock {Unveiling the Mechanisms of Electron Energy Spectrum Evolution}.
\newblock {\em arXiv e-prints}, page arXiv:2412.09016, December 2024.

\bibitem{PhysRevD.109.063027}
Xu-Lin Dong, Yu-Hua Yao, Yi-Qing Guo, and Shu-Wang Cui.
\newblock New understanding of nuclei spectra properties observed by the ams-02
  experiment.
\newblock {\em Phys. Rev. D}, 109:063027, Mar 2024.

\bibitem{2026arXiv260207610N}
Lin {Nie}, Yi-Qing {Guo}, and Si-Ming {Liu}.
\newblock {Two-component $\gamma$-Ray Structure from the CR Sources Within
  Dense Clouds}.
\newblock {\em arXiv e-prints}, page arXiv:2602.07610, February 2026.

\bibitem{Nie:2024txr}
Lin Nie, Yu-Hai Ge, Yi-Qing Guo, and Si-Ming Liu.
\newblock {Geminga: A Window into the Role Played by the Local Halo in the
  Cosmic-Ray Propagation Process}.
\newblock {\em Astrophys. J.}, 991(1):19, 2025.

\bibitem{evoli2008cosmic}
Carmelo Evoli, Daniele Gaggero, Dario Grasso, and Luca Maccione.
\newblock Cosmic ray nuclei, antiprotons and gamma rays in the galaxy: a new
  diffusion model.
\newblock {\em Journal of Cosmology and Astroparticle Physics}, 2008(10):018,
  2008.

\bibitem{gleeson1968solar}
LJ~Gleeson and WI~Axford.
\newblock Solar modulation of galactic cosmic rays.
\newblock {\em Astrophysical Journal, vol. 154, p. 1011}, 154:1011, 1968.

\end{thebibliography}

\end{document}